\documentclass{aa}  
\usepackage{natbib}
\usepackage{graphicx}
\usepackage{txfonts}
\usepackage{hyperref}
\renewcommand{\rm}[1]{\mathrm{#1}}

\begin{document}

   \title{{Radiation hydrodynamics of star-disc collisions\\ for quasi-periodic eruptions}}

   \author{T. Jankovi\v{c} \inst{1,2}
            \and
          C. Bonnerot \inst{3}
        \and
          S. Karpov \inst{1}
          \and
          A. Jurca \inst{2}
          }

   \institute{Institute of Physics of the Czech Academy of Sciences, Na Slovance 1999/2,182 21 Praha 8, Prague, Czech Republic\\
              \email{jankovic@fzu.cz}
        \and
        Centre for Astrophysics and Cosmology, University of Nova Gorica, Vipavska 11c, 5270 Ajdov\v{s}\v{c}ina, Slovenia
      \and
           School of Physics and Astronomy \& Institute for Gravitational Wave Astronomy, University of Birmingham, Birmingham B15 2TT, UK
             }

   \date{Received XX; accepted XX}

  \abstract
   {Quasi-periodic eruptions (QPEs) are recently discovered transients of an unknown nature that occur near supermassive black holes and feature bright X-ray bursts separated by hours to days. A promising model for QPEs is the star-disc collisions model, where a star repeatedly interacts with an accretion disc around a black hole, creating shocks that expel dense outflows of gas from which radiation emerges.} 
   {We investigate the dynamics of star-disc collisions, the properties of the outflows, and the resulting radiation signatures. Our study focuses on the generic case where the star remains unperturbed by the collision and the stellar crossing time through the disc is sufficiently long for shocked gas to flow around the star.}
   {We performed a three-dimensional (3D) radiation-hydrodynamics simulation of the star-disc collision. The star was modelled as a solid, spherical body, and the interaction was simulated for a small, local section of the accretion disc.}
   {We found that star-disc collisions generate a nearly paraboloidal bow shock. The heating of gas is not confined to the column of gas directly ahead of the star but also extends laterally as the shock front expands sideways while travelling with the star. As the star crosses the disc, it injects momentum preferentially along its direction of motion, leading to an asymmetric redistribution of energy and momentum. As a result, two outflows emerge on opposite sides of the disc with different properties: the forward outflow expands faster, contains more mass, carries more energy, and is about twice as luminous as the backward outflow.}  
 {Our findings suggest that the asymmetry in outflow properties and luminosity arises naturally from the collision dynamics, which offers a possible explanation for the alternating 'strong-weak' flare patterns observed in several QPE sources.}


\titlerunning{Radiation-hydrodynamics of star-disc collisions}
   \maketitle
%
%-------------------------------------------------------------------

\section{Introduction}

Quasi-periodic eruptions (QPEs) are a recently discovered class of luminous nuclear transients of an unknown nature that occur near supermassive black holes (SMBHs).  These X-ray sources exhibit bright, hour-long bursts superimposed on a quiescent X-ray emission, with recurrence times of hours to days and reaching peak luminosities of approximately $10^{41}-10^{43}\,\rm{erg\,s^{-1}}$ (e.g. \citealt{Miniutti2019, Giustini2020, Arcodia2021, Arcodia2022, Miniutti2023,Nicholl_2024Natur.634..804N,chakraborty2025discoveryquasiperiodiceruptionstidal}). A significant part of the known QPEs display an alternating 'long-short' or 'strong-weak' pattern between consecutive flares, while the rest exhibit irregular variations of their amplitudes and intervals between them.

The origin of QPEs remains an open question, and several models have been proposed to explain their nature. Suggested mechanisms include accretion disc instabilities (e.g. \citealt{Raj_2021ApJ...909...82R,Sniegowska2023}), gravitational lensing in a binary black hole system \citep{Ingram2021}, mass transfer from a secondary orbiting object (e.g. \citealt{Metzger2022,Wang_2022ApJ...933..225W,Lu_2023MNRAS.524.6247L,Linial_2023ApJ...945...86L}), and Lense-Thirring precession of super-Eddington flows \citep{Middleton_2025MNRAS.537.1688M}. Another class of models considers interactions between a compact object (such as a black hole or a star) and a pre-existing accretion disc (e.g. \citealt{Ivanov_1998ApJ...507..131I,Dai2010,Xian2021,Sukova2021,Krolik_2022ApJ...941...24K,Linial2023, Tagawa_2023MNRAS.526...69T, Zhou_2024PhRvD.110h3019Z,Linial_2024ApJ...973..101L,Yao_2025ApJ...978...91Y,vurm_2025ApJ...983...40V}; {\citealt{Lam_2025PhRvD.112h3006L}}).

\begin{figure*}[h] %  figure placement: here, top, bottom, or page
   \centering
   \includegraphics[width=\textwidth]{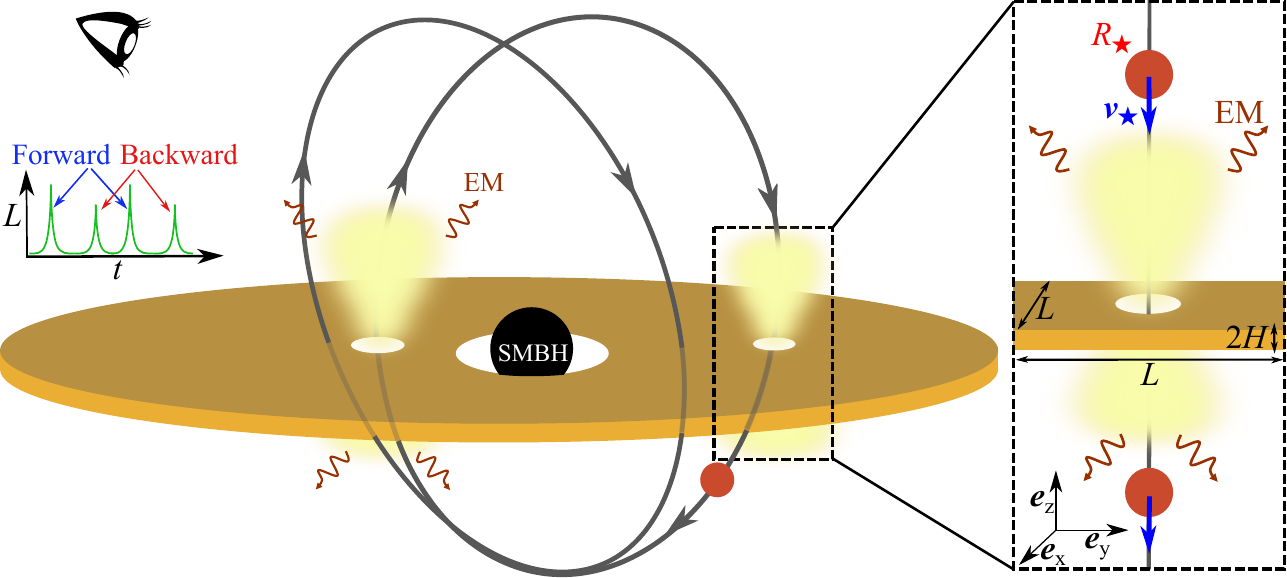}
   \caption{Schematic illustration of the star-disc collisions model for QPEs. The red circle, orange ellipse, and black circle represent the star, the accretion disc, and SMBH, respectively. The star follows an orbit represented by the grey ellipses, intersecting the accretion disc twice per orbital period. Each collision generates a flare of electromagnetic (EM) radiation, resulting in a luminosity, $L,$ seen by a distant observer, as illustrated in the graph on the left-hand side. If one of the outflows is more luminous, the observer would see a recurrent pattern of brighter and dimmer flares. The inset on the right zooms into the collision region, {marked by a dashed rectangle,} and highlights the initial setup of our simulations: a localized section of the disc with horizontal and vertical extents of $L$ and $2H$, respectively, and a star, with radius $R_\star$, moving perpendicularly towards the disc with velocity $v_\star$.}
        \label{fig:qpe_sketch}
\end{figure*}

Among the most promising candidates is the star-disc collisions model, which is illustrated in Figure \ref{fig:qpe_sketch}. In this model, a star is brought into the proximity of a SMBH {due to two-body scattering and gravitational wave radiation}, and an accretion disc surrounds the SMBH. Because the star's orbit is inclined relative to the disc midplane, it intersects the disc twice per orbit, generating shocks that expel dense, optically thick gas clouds above and below the disc. These clouds expand due to radiation pressure, and over timescales of hours their density drops sufficiently for radiation to escape. This model naturally explains the alternating 'long-short' patterns of QPEs due to a moderately eccentric orbit of the star. The observed 'strong-weak' pattern could arise from the asymmetry in the ejecta produced on either side of the disc following the star's passage. Consequently, one collision would appear brighter and the other dimmer to an observer, depending on the viewing angle. The locations on the disc where star-disc collisions occur could vary with time due to Lense-Thirring precession of the accretion disc (e.g. \citealt{Franchini_10.1093/mnras/stv2417,Chakraborty_2024}) or due to general relativistic effects on the stellar orbit (e.g. \citealt{Franchini2023}), changing both the stellar velocity and the disc properties at the collision region. Consequently, strong general relativistic effects could explain the irregular class of QPEs, characterized by the erratic 'long-short' and 'strong-weak' patterns observed between consecutive flares.

{Previous studies have explored important aspects of the star--disc collision scenario, although many did not address the physical regime most relevant for QPEs. \citet{Syer1991} proposed that repeated star--disc encounters can grind stars into short-period orbits, which has implications for fueling of active galactic nuclei (AGNs). \citet{Zurek_1994ApJ...434...46Z} investigated hypersonic star--disc collisions and argued that such encounters can generate downstream wakes and extended gaseous tails, with turbulence being potentially important for their subsequent evolution. \citet{Armitage_1996ApJ...470..237A} performed smoothed-particle hydrodynamics (SPH) simulations of red giant--disc collisions, and studied stripping of the stellar envelope and its possible contribution to AGN fueling. \citet{Subr_1999A&A...352..452S} further examined the long-term orbital evolution of repeatedly colliding orbiters, and derived terminal radii and alignment timescales for eccentric trajectories that intersect the disc once or twice per orbit.} \citet{Dai2010} and \citet{Sukova2021} performed numerical simulations of star-disc interactions but did not include radiation transport or the unique parameter space associated with QPEs. Analytical studies (e.g. \citet{Nayakshin2004}, \citet{Xian2021}, \citet{Franchini2023}, and \citet{Tagawa_2023MNRAS.526...69T}), have calculated collision locations and derived light curve predictions, but they lack the detailed treatment of gas-radiation coupling necessary to fully characterize the observational features.

More recent works have investigated star-disc collisions in greater detail using hydrodynamical and radiation-hydrodynamical simulations.  \citet{Yao_2025ApJ...978...91Y} performed hydrodynamics simulations of repeated star-disc collisions that focused on the effect of the disc on the star over multiple collisions. Their work characterizes the effect of collisions on the star's structure, including mass stripping, and estimates the QPE lifetime. \citet{vurm_2025ApJ...983...40V} performed one-dimensional (1D) Monte Carlo radiation-hydrodynamics simulations and focused on the star-disc collision scenarios for thin discs, through which the star passes faster than the shocked gas can flow around the star. They investigated the photon production and thermalization within radiative shocks. They also constrain the parameter space required to reproduce the observed properties of QPEs. \citet{Huang_2025ApJ...993..186H}  performed two-dimensional (2D), multi-group radiation-hydrodynamic simulations and focused on the multi-band emission signatures from a solar-like star colliding with an accretion disc. They find that the collision produces optically thick ejecta on both sides of the disc, with asymmetric properties arising from asymmetric shock breakout dynamics. The bolometric light curves, rise and decay timescales, and spectra from the resulting outflows are broadly consistent with the observed properties of short-period QPEs. Through 2D
and 3D hydrodynamical simulations, {\citet{Huang_2026arXiv260400953H} show  that more oblique collisions, which arise naturally due to disc rotation, lead to a smaller luminosity asymmetry between the forward and backward ejecta than perpendicular collisions.} {\citet{Liu_2026arXiv260300226L} performed global 3D hydrodynamical simulations comparing star--disc and stellar mass black-hole--disc collisions. They find that collisions with black holes produce more symmetric forward and backward ejecta than collisions with stars, which potentially leads to more similar forward and backward flares.}

In this paper, we study the dynamics of star-disc collisions and the properties of the emerging radiation by performing the first 3D radiation-hydrodynamics simulations of star-disc collisions. {We focus on the regime relevant to most QPEs, where the disc's vertical extent is a few times greater than the stellar size, allowing the shocked gas to flow around the star while leaving the star largely unperturbed by the interaction.} We analyse the redistribution of the energy and momentum injected by the star into the disc and the properties of ejected outflows. {We evaluate the emerging radiation from the outflows and potentially detectable signatures of such collisions.}

This paper is structured as follows. In Section \ref{sec:method}, we describe the physical conditions of the star-disc collisions model, the numerical setup, and the initial conditions. In Section \ref{sec:results}, we present the results of our simulations, including the shock structure, outflow properties, and resulting light curves. In Section \ref{sec:discussion}, we discuss the implications of our findings and compare them to observed QPE properties. We summarize our main conclusions in Section \ref{sec:conclusion}.

%-------------------------------------------------------------------
\section{Methodology}
\label{sec:method}

\subsection{Initial conditions} \label{subsec:star_disc}

We performed a 3D radiation-hydrodynamics simulation of the star-disc collision using the SPH code \textsc{Phantom} \citep{Price2018} in a reference frame co-moving with the accretion disc. The simulation setup is illustrated on the right side of Figure \ref{fig:qpe_sketch}, marked by a dashed black rectangle. The coordinate system is defined by unit vectors $\boldsymbol{e}_\rm{x}$, $\boldsymbol{e}_\rm{y}$, $\boldsymbol{e}_\rm{z}$, and the origin is placed at the disc centre. Initially, the centre of mass (CoM) of the star with radius $R_\star$ was positioned at $(0,0,5R_\star)$, moving perpendicularly toward the disc with velocity $\boldsymbol{v}_\star=(0,0,-v_\star)$. This setup corresponds to a simplified situation in which the star and disc have no relative azimuthal motion, although in reality such motion is present. We discuss this further in Section \ref{subsec:effect_params} and defer a more detailed study to future work. The disc density was assumed to be uniform and equal to $\rho_\rm{d}$. {This simplified vertical structure was adopted as a controlled fiducial setup, allowing us to investigate the dynamics of the star--disc collision, the properties of the outflows, and the resulting radiation signatures in the simplest case before introducing additional complexities. The effect of more realistic vertical stratification is deferred to follow-up work.}

The key system parameters were chosen according to \citet{Linial2023}, assuming an SMBH mass $M_\mathrm{bh}=10^6\,M_\odot$ and QPE period $P_\mathrm{QPE}=4\,\mathrm{h}$. The disc properties were obtained assuming a steadily accreting, optically thick, radiation pressure-dominated $\alpha$-disc \citep{Shakura1973}. Since this is extensively described in \citet{Linial2023}, we focus only on describing our numerical setup. Specifically, we set $R_\star=R_\odot$, $v_\star = 0.1c$, {disc height $H=3R_\star$, and disc density} $\rho_\rm{d} = 3.7 \times 10^{-8}\,\mathrm{g\,cm^{-3}}$, where $c$ is the speed of light. We have chosen $\rho_\rm{d}$ such that the mass of gas in the cylindrical column directly intercepted by the star was $M_\rm{d}=1.2\times10^{-7}\,M_\odot$.\footnote{$M_\rm{d}$ is half the value given by Equation 13 in \citet{Linial2023}, because we perform our simulation in a reference frame co-moving with the disc. Therefore, we do not consider the additional mass swept up by the star due to the relative transverse motion of the disc into the star's path.} To prevent significant expansion of the disc due to {the internal pressure within the disc medium} over the course of the simulation, we set the specific internal energy to $u_\rm{d} = 10^{-5} v_\star^2$. We discuss the effect of different key system parameters in Section \ref{subsec:effect_params}. 

To reduce computational costs, we simulated only a localized region of the accretion disc. The disc was constructed by randomly distributing $1.5\times 10^6$ SPH particles in a cuboidal domain centred at the origin, with horizontal dimensions $L=12R_\star$ and vertical extent $2H$. The particles were distributed according to a random uniform probability density function. SPH particle's mass was chosen to achieve a target density equal to $\rho_\rm{d}$. To smooth the initial particle distribution, we performed a relaxation process by evolving the system under periodic boundary conditions.

The accretion disc in a realistic system exhibits differential rotation, meaning that gas at different radial distances $R$ from the SMBH orbits at different angular velocities. However, in our simulation, we neglect this effect. To assess the validity of this approximation, we compare the disc crossing timescale $t_\mathrm{cr} \approx H / v_\star$ to the local shearing timescale $t_\mathrm{shear} \approx \Omega ^{-1}$, where $\Omega = v_\star / R$ is the Keplerian angular velocity. This yields a ratio ${t_\mathrm{cr}}/{t_\mathrm{shear}} \approx H/R=0.02 \ll 1$,
using $R=100R_\rm{g}$, corresponding to a radius of a circular stellar orbit with orbital period $\approx 8\,$h,  and value of $H$ from our simulation. This indicates that differential rotation has little time to affect the shock structure or outflow geometry. However, disc shearing could become relevant in scenarios where $t_\mathrm{cr}$ is longer, such as for thicker discs or more inclined stellar orbits.

\subsection{SPH Simulations}\label{subsec:simulations}

The gas pressure was modelled with an adiabatic equation of state with the adiabatic index \(\Gamma=5/3\), and radiation diffusion was treated in the flux-limited diffusion approximation (e.g. \citealt{Whitehouse_2004MNRAS.353.1078W, Whitehouse_2005MNRAS.364.1367W, Bate_2015MNRAS.449.2643B, Bonnerot2021}; {\citealt{Lau_2025A&A...699A.274L}}), which assumes local thermodynamical equilibrium (LTE).\footnote{Similar to supernova shock breakouts (e.g. \citealt{Weaver_1976ApJS...32..233W}), the gas and radiation in the expanding shocked disc material may not be in LTE. Due to the low densities and rapid expansion of the outflow, photon production can be inefficient, leading to spectra with higher effective temperatures than expected for blackbody emission (e.g. \citealt{Nakar_2010, vurm_2025ApJ...983...40V}). We discuss this further in Section \ref{sec:limit}.}. We note that the version of \textsc{Phantom} used in this research does not include a surrounding low-density medium into which radiation can freely escape once the gas becomes optically thin. This could lead to an unphysical accumulation of radiation in this regime, which could affect the results of our simulations. We investigate this in Appendix \ref{app:rad_accum} and conclude that this effect has negligible consequences on our results.

The radiation diffusion flux $F_\mathrm{diff}$ was calculated as
\begin{equation}\label{eq:Fdiff}
    \boldsymbol{F}_\mathrm{diff}=-\frac{\lambda c}{\kappa_\rm{s}} \frac{\boldsymbol{\nabla} e_\rm{rad}}{\rho},
\end{equation}
where $\rho$, $e_\rm{rad}$, and $\kappa_\rm{s}=0.34\,\rm{cm^2g^{-1}}$, are the gas density, radiation energy density, and electron scattering opacity, respectively. The flux limiter $\lambda$ prevents the unphysical and arbitrarily fast transport of radiation in an optically thin gas. \textsc{Phantom} adopts the prescription by \citet{Levermore_1981ApJ...248..321L}, where the flux limiter is given by $\lambda = (2 + K)/(6 + 3K + K^2)$ 
for $\lambda \leq 1/3$, where $K = |\nabla e_\rm{rad}| / (\kappa_\rm{s} \rho e_\rm{rad})$.

In the star--disc collision model, the prompt shock formation and outflow launching occur on a timescale much shorter than the local orbital period. We therefore neglect the SMBH's gravitational field, as well as Coriolis and centrifugal forces.\footnote{The characteristic timescale associated with the vertical component of the SMBH gravity is $t_z \sim 2\pi \Omega^{-1}\sim8\,$h, assuming $v_\star=0.1c$, $R=100R_{\rm g}$, and $M_{\rm bh}=10^6\,M_\odot$, which is much longer than the $\sim 10$ min duration of our simulation. Thus, the vertical component of the SMBH gravity has little effect on the early gasdynamics studied here. {However, we note that disc self-gravity can modify the pre-collision vertical structure of the accretion disc, affecting both its geometrical thickness and density distribution. In such cases, the disc may deviate from the non-self-gravitating thin-disc model, and a different disc model would be required (e.g. \citealt{Hure_2000A&A...358..378H,Gangardt_2024MNRAS.530.3689G}). We defer the investigation of such models to future work.}} The self-gravity of the gas and the gravity of the star were also not considered, as the kinetic energy imparted to the shocked gas is significantly higher than these gravitational energies on the timescales relevant to the prompt shock dynamics and outflow formation (e.g. {\citealt{Huang_2025ApJ...993..186H,Huang_2026arXiv260400953H}}). {We also do not include a pre-existing coronal or ambient medium above the disc, since a bright AGN-like hot X-ray corona appears disfavoured in QPE sources, whose quiescent emission is typically weak and soft in X-rays. A weaker disc atmosphere or warm coronal layer would be much less dense than the collision ejecta and is therefore not expected to significantly affect the early gasdynamics considered here.} We model the star as a rigid sphere with radius \(R_\star\), where its interaction with the gas is treated as elastic collisions with individual SPH particles. Specifically, when a gas particle reaches the stellar surface, its velocity component normal to the surface is reversed, while the tangential component is preserved. Additionally, we neglect any losses of the stellar momentum due to these collisions by keeping the stellar velocity fixed, which is justified by $M_\star$ being much larger than the intercepted disc mass $M_\rm{d}$, with $M_\star/M_\rm{d}\approx 10^{7}$ assuming $M_\star = M_\odot$. {We also neglect the effect of the star's own radiation pressure and wind on the disc medium}.\footnote{To assess this we compare the pressure due to radiation momentum flux at the stellar surface $P_\mathrm{rad,\star}\sim L_\star/(4\pi R_\star^2 c)$, and wind ram pressure $P_\mathrm{ w,\star}\sim \dot{M}_{\rm w}v_{\rm w}/(4\pi R_\star^2)$, with the disc ram pressure $P_\mathrm{ ram,d}\sim \rho_{\rm d}v_\star^2$, where $L_\star$, $\dot{M}_{\rm w}$, and $v_{\rm w}$ are the stellar luminosity, the wind mass-loss rate, and the wind velocity, respectively. For a Sun-like star with $L_\star=L_\odot$, $R_\star=R_\odot$, $\dot{M}_{\rm w}=2\times10^{-14}\,M_\odot\,\mathrm{ yr^{-1}}$, and $v_{\rm w}=500~\mathrm{ km\,s^{-1}}$, we obtain $P_\mathrm{rad,\star}/P_\mathrm{ ram,d}\sim 10^{-12}$ and $P_\mathrm{w,\star}/P_\mathrm{ ram,d}\sim 10^{-15}$. Even for an extreme case of a Wolf--Rayet star with $L_\star=3\times10^5\,L_\odot$, $R_\star=3\,R_\odot$, $\dot{M}_{\rm w}=10^{-5}\,M_\odot\,\mathrm{ yr^{-1}}$, and $v_{\rm w}=2000~\mathrm{ km\,s^{-1}}$, one finds $P_\mathrm{ rad,\star}/P_\mathrm{ ram,d}\sim 10^{-7}$ and $P_\mathrm{ w,\star}/P_\mathrm{ ram,d}\sim 10^{-7}$. Moreover, we do not expect the star's own wind or radiation pressure to significantly affect the subsequent outflow evolution, since the ejecta expand with characteristic velocities $v_\star\sim 0.1c$, far exceeding even strong stellar-wind velocities.}

The initial star-disc configuration described in Section \ref{subsec:star_disc} resulted in a smoothing length of \(h_\mathrm{sl} \approx (m / \rho)^{1/3}\approx 0.05R_\star\), where $m$ is particle mass. This ensured that $h_\mathrm{sl}$ remained much smaller than the size of the star and disc, which provided sufficient resolution to capture the collision dynamics accurately. We verified this by performing simulations with $h_\mathrm{sl} \approx 0.01R_\star$ and $h_\mathrm{sl}\approx 0.005R_\star$ and found no qualitative differences.

\section{Results}
\label{sec:results}

\subsection{Collision dynamics}\label{subsec:coll_dyn}

\begin{figure*}[h] %  figure placement: here, top, bottom, or page
   \centering
   \includegraphics[width=\textwidth]{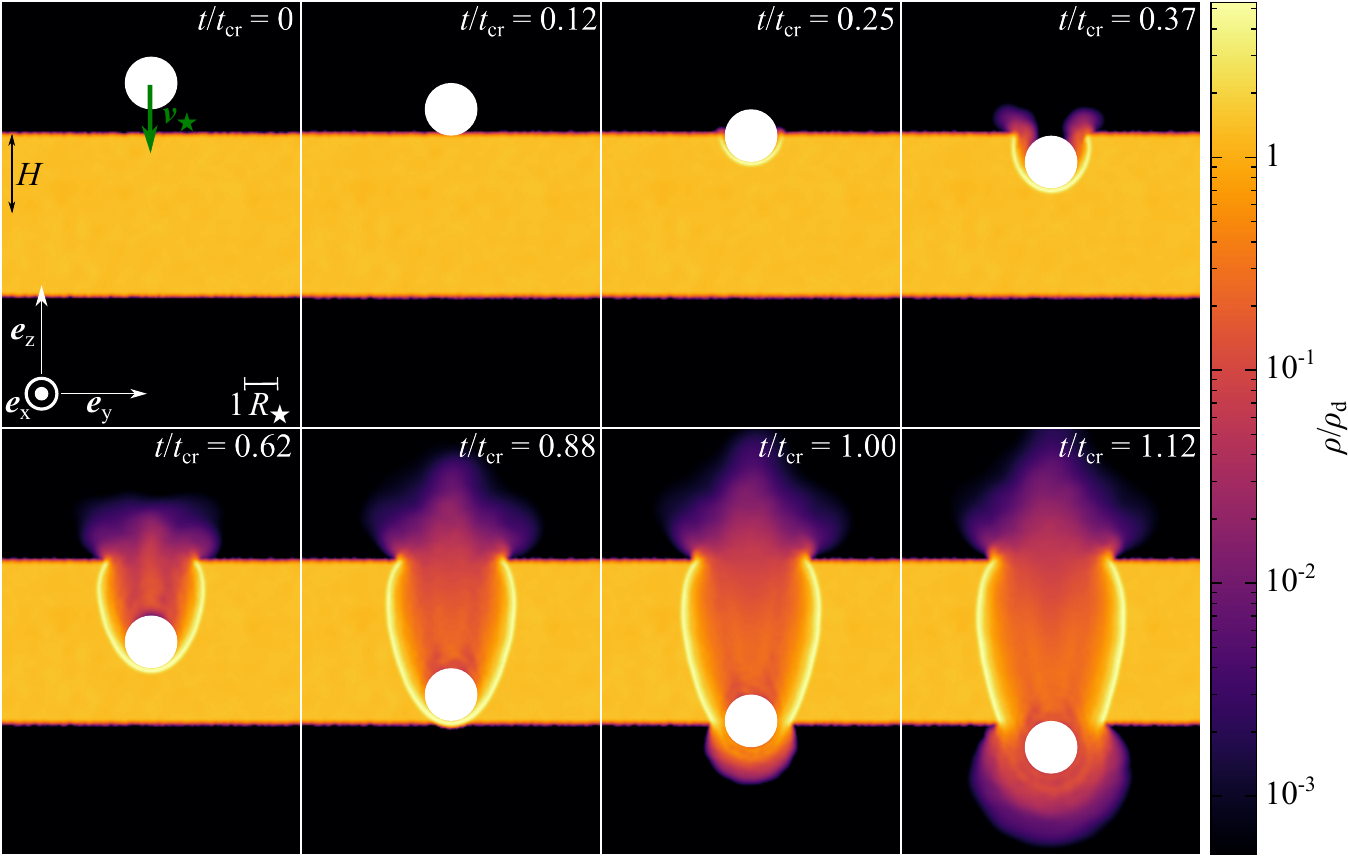}
   \caption{Gas density slices in the $yz$-plane at $x=0$ at different times. The white circle denotes the star.}
        \label{fig:qpe_sim_density}
\end{figure*}

\begin{figure*}[h] %  figure placement: here, top, bottom, or page
   \centering
   \includegraphics[width=\textwidth]{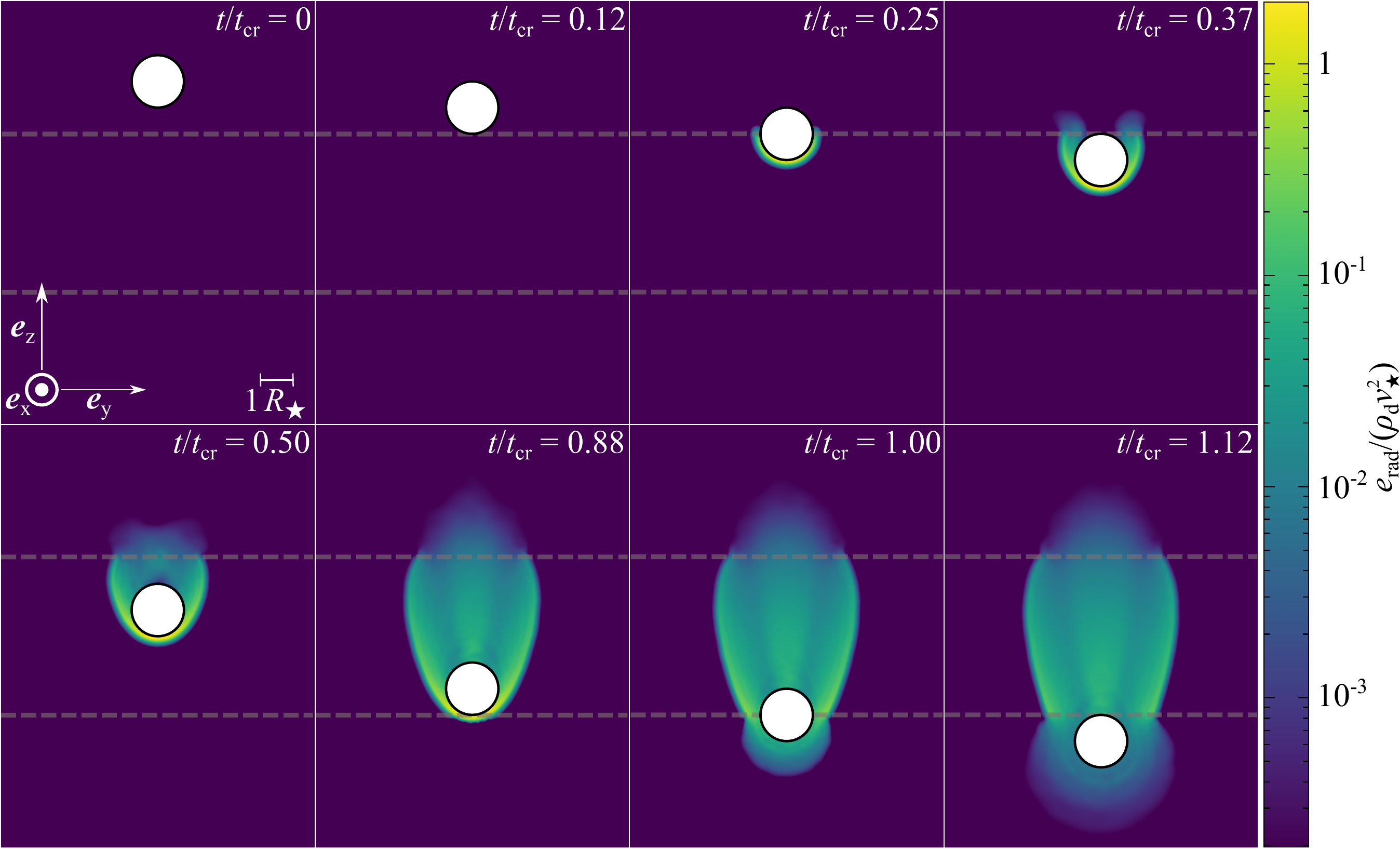}
   \caption{Radiation energy density slices in the $yz$-plane at $x=0$ at different times. The white circle denotes the star, while the dashed grey lines mark the outer edges of the accretion disc.}
        \label{fig:qpe_sim_erad}
\end{figure*}

The dynamics of the star-disc interaction unfold in several stages, driven by the formation of shocks, the redistribution of momentum and energy, and the subsequent ejection of gas. Figures \ref{fig:qpe_sim_density} and \ref{fig:qpe_sim_erad} show $\rho$ and $e_\rm{rad}$, respectively, contained in slices in the $yz$-plane at $x=0$. We normalize time to the crossing time of the star through the disc $t_\rm{cr}=2(H+R_\star)/v_\star$.

At \(t/t_\rm{cr} \approx 0.12\), the star collides supersonically with the disc, forming a radiation-pressure-dominated bow shock. This leads to a sharp increase in \(\rho\) and temperature, enhancing photon production and increasing \(e_\mathrm{rad}\) (see panel at \(t/t_\rm{cr} \approx 0.12\) in Figure \ref{fig:qpe_sim_erad}). The flow of gas around the star is governed by post-shock velocities, ranging between \(\sim \left (\Gamma-1 \right) \left (\Gamma+1 \right)^{-1} v_\star = v_\star / 7\) (for a radiation-pressure dominated gas with \(\Gamma = 4/3\)) and the post-shock sound speed \(c_\mathrm{s} \approx v_\star\). 

By \(t/t_\rm{cr} \approx 0.25\), a quasi-steady state is established where the inflow of material into the shock front is balanced by the outflow of gas escaping around the star (see Appendix \ref{app:Mcap}). The shock front reaches the upper disc boundary, triggering a shock breakout (see e.g. \citealt{Linial_2019PhFl...31i7102L}), where trapped radiation escapes as the post-shock gas expands into a low-pressure region. Gas flowing around the star converges in its wake, forming a secondary shock (see panel at \(t/t_\rm{cr} = 0.5\)). This process re-heats the gas and redistributes its momentum and energy, forming a distinct ejecta component in the outflow from the upper disc edge visible in Figure \ref{fig:qpe_sim_density} as the vertical region with an increased density behind the star at \(t/t_\rm{cr} \gtrsim 0.5\). 

At $t/t_\rm{cr} \approx 0.88$, the bow shock reaches the lower disc boundary, triggering a shock breakout and launching a nearly spherical outflow along the forward direction. As the star begins to emerge from the disc (see panel at $t/t_\rm{cr} = 1$ in Figure \ref{fig:qpe_sim_density}), gas starts to flow outward from the lower disc edge, in addition to that produced directly by the bow shock (see Section \ref{subsec:outflow} for a more detailed explanation). The ejecta from the upper and lower disc boundaries are asymmetric, as seen at $t/t_\rm{cr}=1.12$.

\begin{figure}
        \centering  
        \begin{minipage}[b]{\linewidth}
        \includegraphics[width=\textwidth]{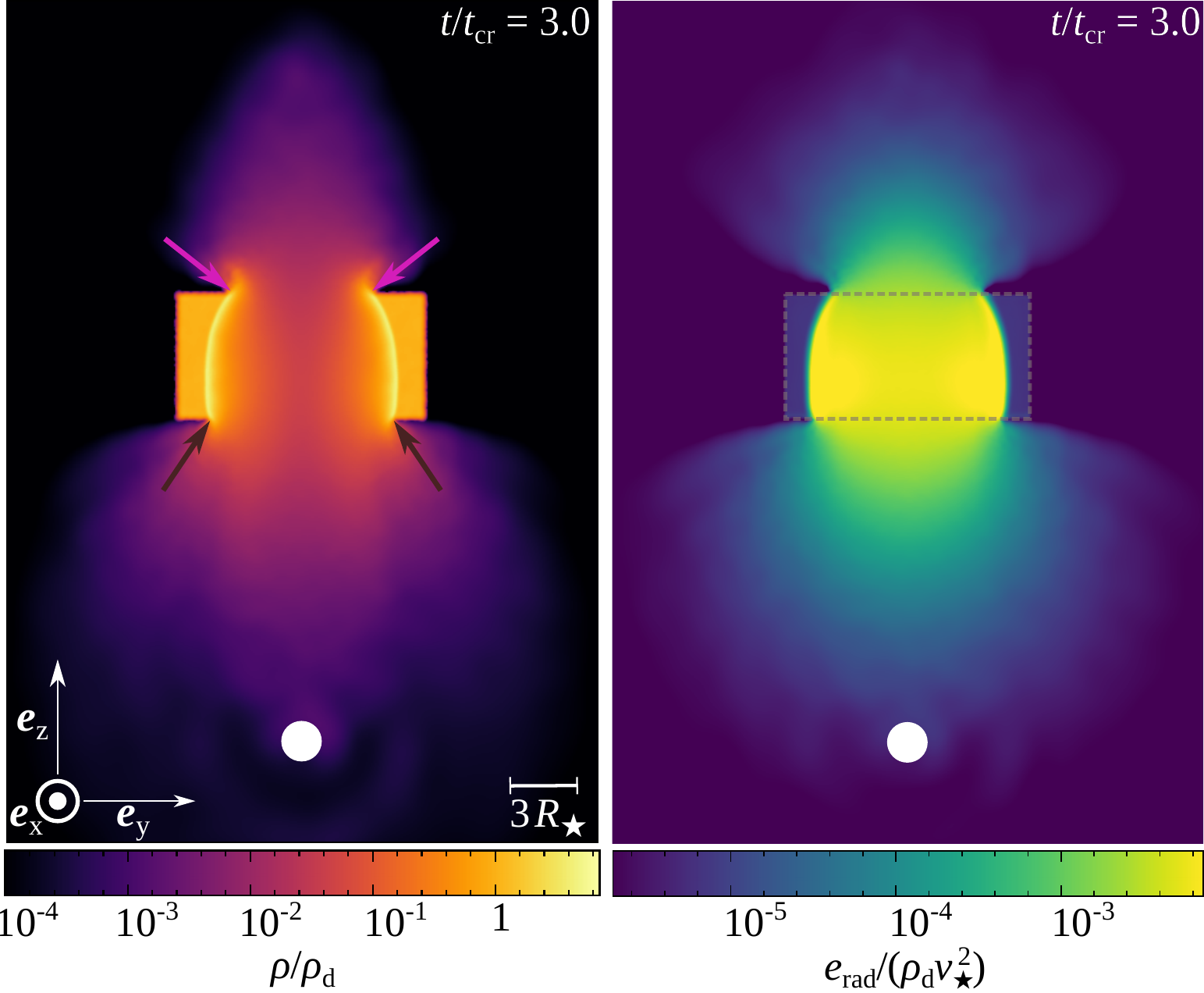}
        \end{minipage} 
        \caption{Gas density (left panel) and radiation energy density (right panel) slices in the $yz$-plane at $x=0$ at $t/t_\rm{cr}=3$. The white circle denotes the star, while the dashed grey lines on the right panel mark the outer edges of the accretion disc. The pink and brown arrow lines in the left panel indicate the lateral shock front distance, $r_\rm{sh}^\perp$.}
        \label{fig:outflow_late}
\end{figure}

In Figure \ref{fig:outflow_late} we show $\rho$ and $e_\rm{rad}$ slices at $t/t_\rm{cr} = 3$. The asymmetry between the outflows is even more apparent at this stage. {The area from which the outflow emerges on the bottom side of the disc is larger than on the upper side.} Additionally, the outflow along the forward direction has a larger mass, higher radiation energy, and expands with higher velocities. We address this asymmetry further in Sections \ref{subsec:outflow} and \ref{subsec:lcs}.

\subsection{Bow shock formation and evolution}\label{subsec:shock}

\begin{figure}
        \centering  
        \begin{minipage}[b]{\linewidth}
        \includegraphics[width=\textwidth]{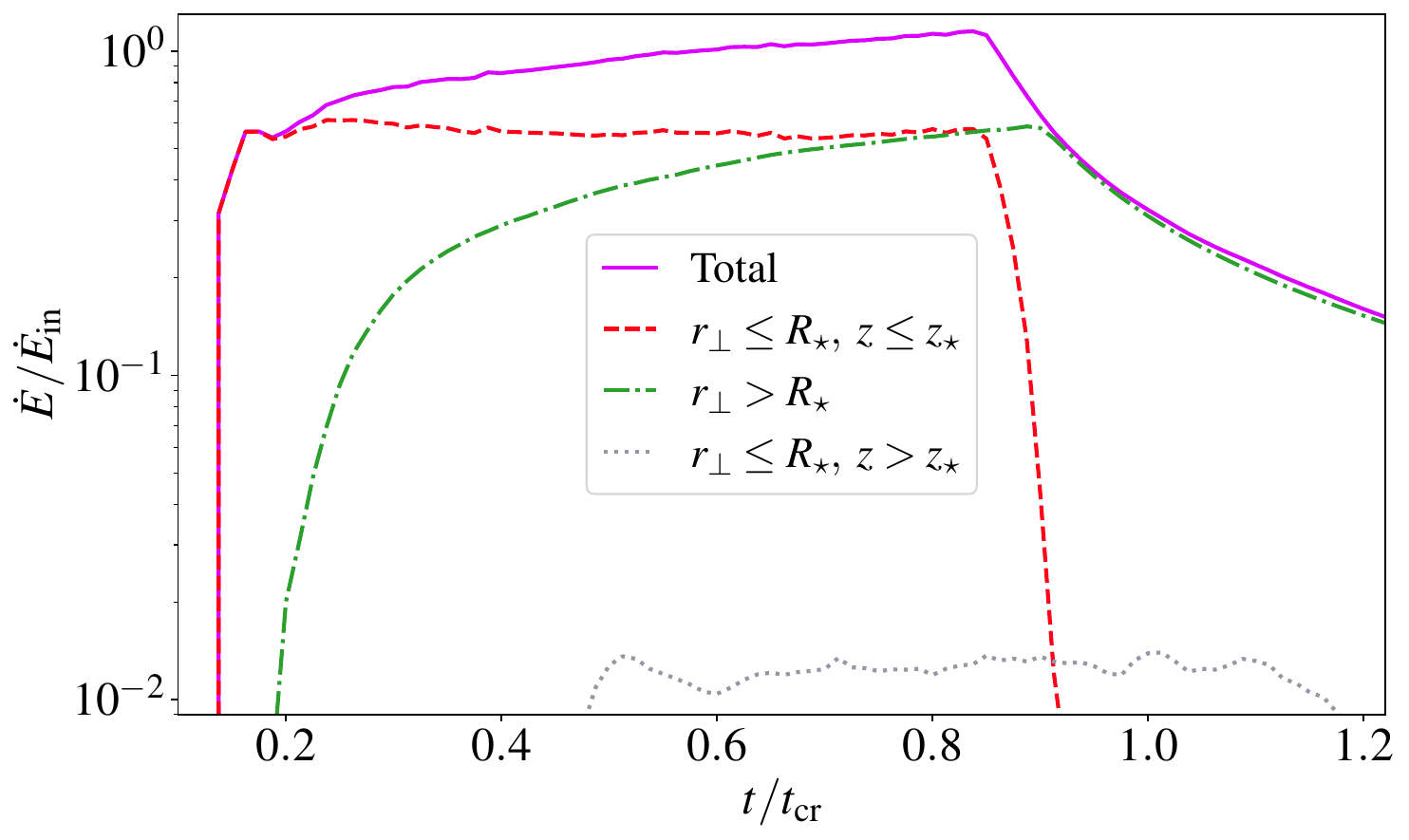}
        \end{minipage} 
        \caption{Shock heating rate, $\dot{E}$ , as a function of time. The solid blue line represents the total $\dot{E}$. Other lines correspond to $\dot{E}$ calculated for gas inside different regions relative to the vertical and lateral distance from the CoM of the star, located at $z_\star$ at time, $t$: gas inside a cylindrical column directly ahead of the star with ${r}_\perp\leq R_\star$ and $z\leq z_\star$ (dash-dotted green), gas outside of the cylindrical column intercepted by the star with ${r}_\perp> R_\star$ (dotted red), and gas inside a cylindrical column directly behind the star with ${r}_\perp\leq R_\star$ and $z> z_\star$ (double-dotted dashed magenta). The coloured vertical lines correspond to specific stages of star-disc collisions (see Figures \ref{fig:qpe_sim_density} and \ref{fig:qpe_sim_erad}).}
        \label{fig:dotE}
\end{figure}

We focus on the formation of the bow shock, which not only travels with the star but also propagates laterally through the surrounding disc material along the direction of the cylindrical radius $\boldsymbol{r}_\perp$ measured from the $z$-axis. Figure \ref{fig:dotE} shows the shock heating rate $\dot{E}$ as a function of time. The values are normalized to the shock heating rate for the gas inside the column through which the star moves  $\dot{E}_\rm{in} =  \dot{M}_\rm{in} \Delta u = 5.6\times 10^{42}\, \mathrm{erg/s}$, where the mass inflow rate into the shock is given by $\dot{M}_\rm{in}= \pi R_\star^2 \rho v_\star = {1.7\times 10^{24}\, \mathrm{g/s}}$, and $\Delta u$ corresponds to the characteristic increase in the specific internal energy of the gas due to the collision. In the rest frame of the star, the velocity of the gas flowing into the shock front is $-\boldsymbol{v}_\star$, meaning that $\Delta u =  {2} { \left({\Gamma} + 1\right)^{-2}} v_\star^2 \approx 0.37 v_\star^2$, obtained from the Rankine-Hugoniot jump conditions assuming a strong shock and ${\Gamma} = 4/3$. 

In Figure \ref{fig:dotE}, the solid magenta line corresponds to $\dot{E}$ calculated for all gas, while the dashed red, dash-dotted green, and dotted grey lines correspond to $\dot{E}$ calculated for gas inside different regions relative to the vertical and lateral distance from the CoM of the star. When the star makes first contact with the disc at $t/t_\rm{cr}\approx 0.12$, the heating rate sharply increases, which is dominated by the direct collision of the star with the disc, corresponding to shock heating in the narrow gas column directly in front of the star (dashed red line). Once the bow shock fully develops, this component stabilizes and remains approximately constant before rapidly decreasing to zero as the star begins to exit the disc at $t/t_\rm{cr}\approx 0.88$. This component remains lower than one, since $\Delta u$ represents the upper limit, which applies to the gas directly in front of the star, where the shock is perpendicular to the flow, and the entire kinetic energy is dissipated. Further from this point, the shock front becomes increasingly oblique, reducing the increase in the specific internal energy of the gas due to the collision since only the velocity component normal to the shock surface is reduced. 

The increase in total $\dot{E}$ (magenta solid line) at $t/t_\rm{cr}\gtrsim 0.25$ results from gas heating at lateral distances $r_\perp \gtrsim R_\star$ (dash-dotted green line), which arises from shock expansion along $\boldsymbol{r}_\perp$. Due to this heating, {the total $\dot{E}/\dot{E}_\rm{in}$ can temporar}ily become larger than one. After the star exits the disc, lateral expansion becomes the dominant source of shock heating, but gradually declines over time. The secondary shock, initiated by the convergence of gas streams in the star's wake, introduces an additional heating component (grey dotted line). This component is delayed compared to the others, reflecting the time required for the gas to flow around the star and converge in the wake. It then remains approximately constant before diminishing after the star exits the disc.

\begin{figure}
        \centering
                \includegraphics[width=\linewidth]{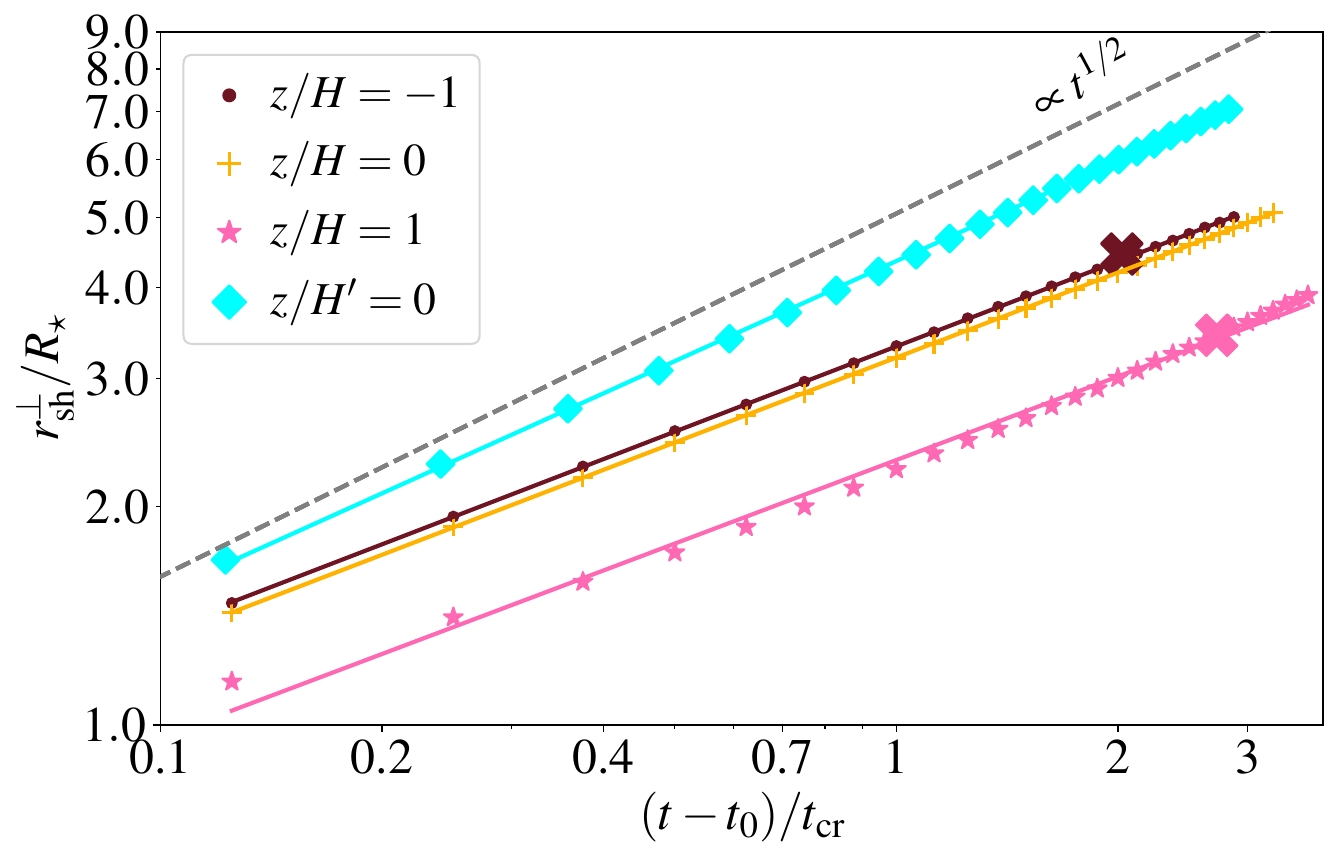}
        \caption{Lateral shock front distance $r_\mathrm{sh}^\perp$ from the axis of symmetry as a function of time. The brown dot ($\bullet$), orange plus ($+$), and pink star ($\star$) symbols correspond to values after the stellar CoM crossed $z/H=-1$, $z=0$, and $z/H=1$, respectively. The cyan diamond ($\blacklozenge$) symbols correspond to values from a simulation with an increased disc height of $H'=50 R_\odot$, which are shifted along the vertical axis for visual clarity. The solid lines represent the best power-law fits to the simulation data. The dashed grey line shows the reference scaling $r_\mathrm{sh}^\perp \propto t^{1/2}$. The enlarged cross ($\boldsymbol{\times}$) symbols indicate $r_\mathrm{sh}^\perp$ near the disc edges at $t/t_\rm{cr}=3$.}
        \label{fig:r_of_t_shock}
\end{figure}

The evolution of the lateral shock front distance $r_\rm{sh}^\perp$ at different vertical positions $z$ is shown in Figure \ref{fig:r_of_t_shock}. We calculate $r_\rm{sh}^\perp$ as the distance from the axis of symmetry where $\rho$ is the highest. We find that the lateral expansion of the shock front approximately follows a power-law scaling  $r_\mathrm{sh}^\perp \propto (t-t_0)^{0.4}$, where $t_0$ corresponds to the time when the tip of the star reaches a given $z$, with only minor variations between different vertical layers --- the lower disc edge (brown '$\bullet$'), midplane (orange '$+$'), and upper boundary (pink '$\star$')  all exhibit a similar scaling. The inferred power-law index is shallower than the asymptotic steady-state scaling $r_\mathrm{sh}^\perp \propto t^{1/2}$ predicted by \citet{Yalinewich_2016ApJ...826..177Y} and \citet{DuPont_2024ApJ...971...34D}.\footnote{\citet{Yalinewich_2016ApJ...826..177Y} proposed a simple analytic argument to obtain this scaling. Consider an object of cross-sectional area $\propto R_\star^2$ moving supersonically at velocity $v_\star$ through a gas of density $\rho$. In each short time interval $\Delta t$, the object injects energy $\propto \rho R_\star^2 v_\star^3 \Delta t$ into the surrounding gas, creating a local explosion that is vertically constrained by neighbouring explosions. This drives a lateral expansion of a disc-like shocked region with radius $r_\rm{sh}^\perp$ and vertical thickness $v_\star \Delta t$, whose energy scales as $\propto \rho r_\perp^2 v_\star\Delta t (r_\perp / t)^2$. Equating injected and stored energy yields $r_\rm{sh}^\perp \propto t^{1/2}$, implying a parabolic shock front shape with $z \propto (r_\rm{sh}^\perp) ^2$.} However, we find that for lower ratios of $R_\star/H$, for which the crossing time $t_\rm{cr}$ is longer, $r_\rm{sh}^\perp$ converges to the expected scaling. This is indicated by cyan '$\blacklozenge$' symbols, which correspond to values obtained from a simulation with an increased disc height of $H'=50R_\star$ and $\rho_\rm{d} = 2.2 \times 10^{-8}\,\mathrm{g\,cm^{-3}}$, while keeping all other parameters fixed, where the choice of $\rho_\rm{d}$ keeps $M_\rm{d}$ the same as for the fiducial simulation. This indicates that in our fiducial simulation, the system is still evolving towards a parabolic steady-state solution.

At late times, $r_\mathrm{sh}^\perp$ near the lower disc edge exceeds that near the upper one, as indicated by the pink and brown arrow lines in the left panel of Figure \ref{fig:outflow_late} and by enlarged '$\boldsymbol{\times}$' symbols in Figure \ref{fig:r_of_t_shock} for $t/t_\mathrm{cr}=3$. Consequently, the lower outflow ejection surface spans a larger area than the upper one. We find that the key factor driving this asymmetry is the stronger lateral detachment of the shock front from the stellar surface on the lower side of the disc. This can be seen in Figure \ref{fig:r_of_t_shock}, where at the earliest times the values of $r_\mathrm{sh}^\perp$ for $z/H=-1$ (brown '$\bullet$') are larger by $\approx 50\%$ compared to those for $z/H=1$ (pink '$\star$'). The shape and detachment of the shock front are predominantly determined by the momentum balance between the incoming and reflected gas relative to the stellar surface, analogous to the scenario of stellar wind bow shocks described by \citet{Wilkin_1996} for example. Initially, in the upper layers of the disc, the reflected momentum is still accumulating, and thus the shock front remains attached to the stellar surface with $r_\mathrm{sh}^\perp \approx 1 R_\star$ (see also panel at $t/t_\rm{cr}=0.25$ in Figure \ref{fig:qpe_sim_density}). As time progresses, the shock front expands outwards. When a quasi-steady state is established, the shape of the front stabilizes, achieving a detached radius of approximately $r_\mathrm{sh}^\perp \approx 1.5 R_\star$ (see also panels at $t/t_\mathrm{cr}=0.50$ and $t/t_\mathrm{cr}=0.88$ in Figure \ref{fig:qpe_sim_density}).

\subsection{Outflow}\label{subsec:outflow}

When the star collides with the disc, it injects energy and momentum into the surrounding gas. These quantities are redistributed in the shocked gas, resulting in the ejection of material along both the backward direction ($+\boldsymbol{e}_\mathrm{z}$; upper disc surface) and the forward direction ($-\boldsymbol{e}_\mathrm{z}$; lower disc surface). 

\begin{figure*}[h]
\centering
\includegraphics[width=0.99\textwidth]{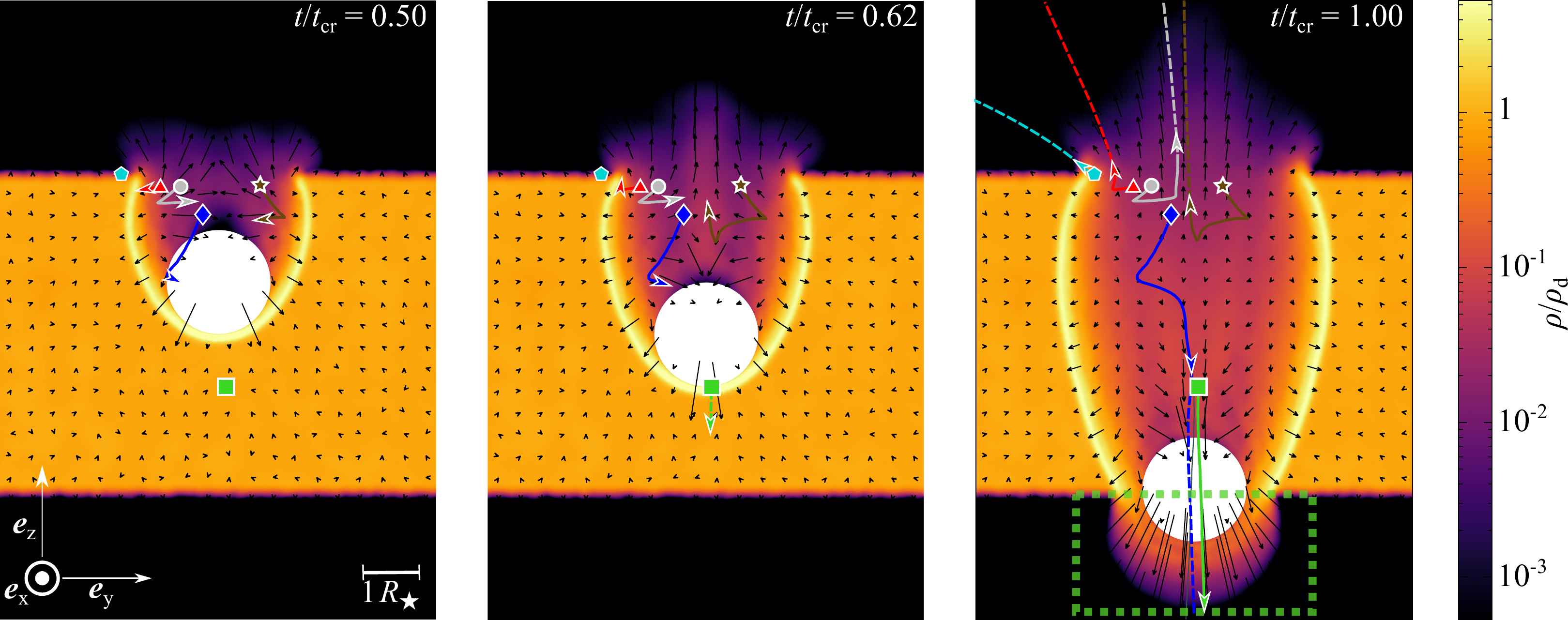}
\caption{Trajectories of six gas particles in the $yz$ plane at $x=0$ at different times plotted over gas density slices (see Figure \ref{fig:qpe_sim_density}). The solid lines indicate the travelled path from the initial particle positions (grey dot `$\bullet$', brown star `$\star$', blue diamond `$\blacklozenge$', green square `$\blacksquare$', red triangle `$\blacktriangle$', and cyan arrow `$\to$' symbols) to their position at the current time (the coloured arrows), while the dashed lines indicate their trajectories at later times. The black arrows show the bulk gas velocities and the white circle denotes the star. The dotted green rectangle indicates the gas ejected during the forward shock breakout.}
\label{fig:qpe_sim_asymm}
\end{figure*}

Figure \ref{fig:qpe_sim_asymm} illustrates the gas dynamics and outflow formation via the trajectories of representative gas particles. The solid lines indicate the travelled path from the initial particle positions to their position at the current time, denoted by the coloured arrows, while the dashed lines indicate their trajectories at later times. At $t/t_\rm{cr} \approx 0.5$ (left panel in Figure \ref{fig:qpe_sim_asymm}), a secondary shock forms behind the star as gas initially flowing around the star converges near the symmetry axis. Two gas particles (grey '$\bullet$' and brown '$\star$' symbols) highlight this process. As these particles are pushed downwards by the star, they are deflected horizontally around it. Then they move towards the low $\rho$ region left in the wake of the star and collide near the symmetry axis (middle panel in Figure \ref{fig:qpe_sim_asymm}). This collision redistributes momentum and energy, effectively creating a velocity reversal region where $v_\rm{z}$ reverses sign (note the change in the vertical direction of the black arrows near the axis of symmetry in the middle panel of Figure \ref{fig:qpe_sim_asymm}). This is a consequence of the radiation pressure force behind the star acting along $+\boldsymbol{e}_\rm{z}$, which can reverse the downward motion of the gas. Thus, both particles ultimately become a part of the backward outflow (right panel in Figure \ref{fig:qpe_sim_asymm}). 

We find that gas originating closer to the axis of symmetry is more likely to be ejected along the forward direction. This is illustrated by the trajectories of two additional gas particles (blue '$\blacklozenge$' and green '$\blacksquare$' symbols) in Figure \ref{fig:qpe_sim_asymm}. Owing to its location closer to the axis of symmetry, the blue '$\blacklozenge$' particle follows a trajectory that takes longer to flow around the star compared to the grey '$\bullet$' and brown '$\star$' particles. By the time it converges behind the star and experiences the secondary shock, it is already well below the velocity reversal region. Due to this, it conserves its downward momentum direction and joins the forward outflow along $-\boldsymbol{e}_\mathrm{z}$. The green '$\blacksquare$' particle does not have time to flow around the star because of its proximity to the axis of symmetry and location closer to the lower edge of the disc. Instead, it remains within the shock cap and is rapidly expelled during the shock breakout.

\begin{figure}
        \centering
    \includegraphics[width=0.99\linewidth]{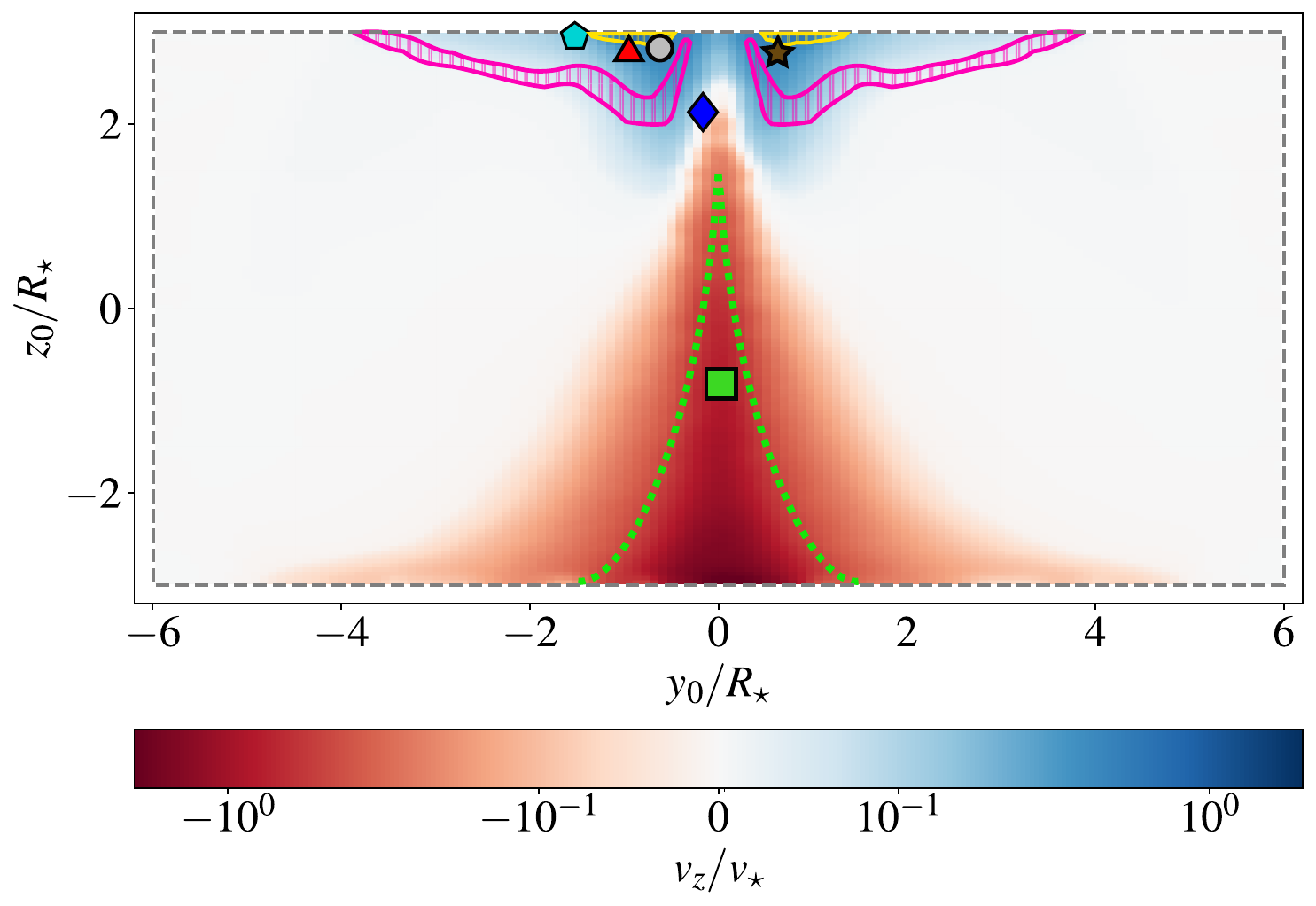}
    \caption{Map of the final vertical velocity, $ v_\rm{z}$ , at $t/t_\rm{cr}=3.8$ for gas shown at its initial position in the $yz$-plane at $t=0$, selected from a slice $|x|/R_\star\leq 0.2$. Gas particles were binned according to their initial positions, and each bin is coloured by the average final $v_\rm{z}$ of the particles it contains. The dotted green region indicates the gas ejected during the forward shock breakout (see the green rectangle in the right panel of Figure \ref{fig:qpe_sim_asymm}). The orange and magenta contours indicate the gas that gets ejected at $t/t_\rm{cr}\approx0.4$ and $t/t_\rm{cr}\approx 3.5$, respectively.}
        \label{fig:vz_vs_r}
\end{figure}

In Figure \ref{fig:vz_vs_r}, we show the vertical velocity distribution of the gas by mapping the final vertical velocity $v_\mathrm{z}$ at $t/t_\mathrm{cr}=3.8$ onto initial gas particle positions in the $yz$-plane at $t=0$. Gas with $v_\mathrm{z}/v_\star>0$ (blue) is ejected along the backward direction, while gas with $v_\mathrm{z}/v_\star<0$ (red) is ejected forwards. Gas initially positioned close to the upper edge of the disc, such as the grey '$\bullet$' and brown '$\star$' particles from Figure \ref{fig:qpe_sim_asymm}, is predominantly ejected backwards. However, gas near the axis of symmetry instead joins the forward outflow because it takes longer to flow around the star (e.g. blue '$\blacklozenge$' particle in Figure \ref{fig:qpe_sim_asymm}). Shocked gas that requires more time to flow around the star than the star needs to leave the disc will be expelled directly downwards during the forward shock breakout. This gas is originally located within the dashed green region in Figure \ref{fig:vz_vs_r} and follows trajectories similar to the green '$\blacksquare$' particle in Figure \ref{fig:qpe_sim_asymm}. Gas farther from the symmetry axis typically achieves lower final $|v_\mathrm{z}|/v_\star$. This is due to the decrease in energy injection by the oblique shock front with increasing distance from the tip of the star.

\begin{figure}[h]
\centering
\includegraphics[width=0.99\linewidth]{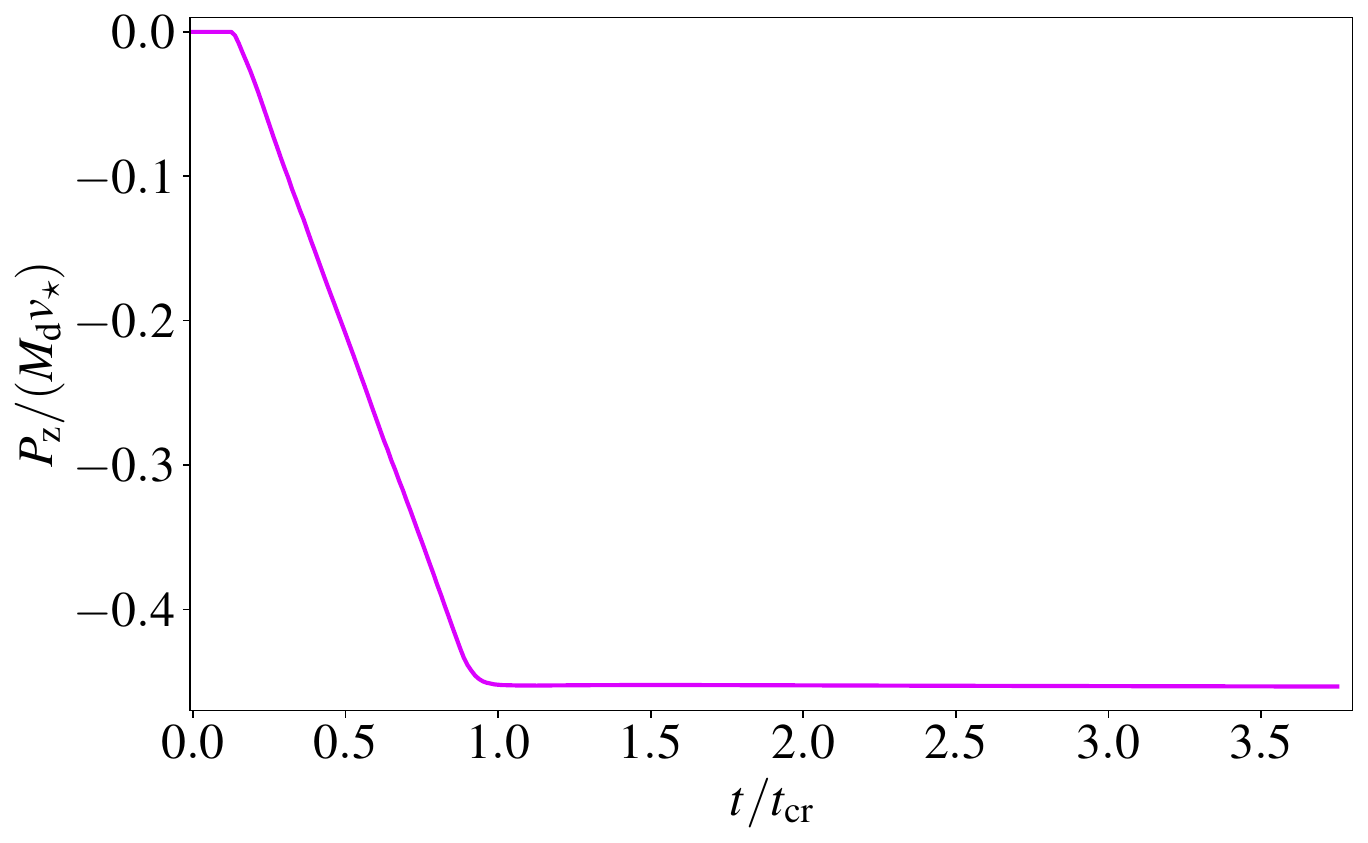}
\includegraphics[width=0.99\linewidth]{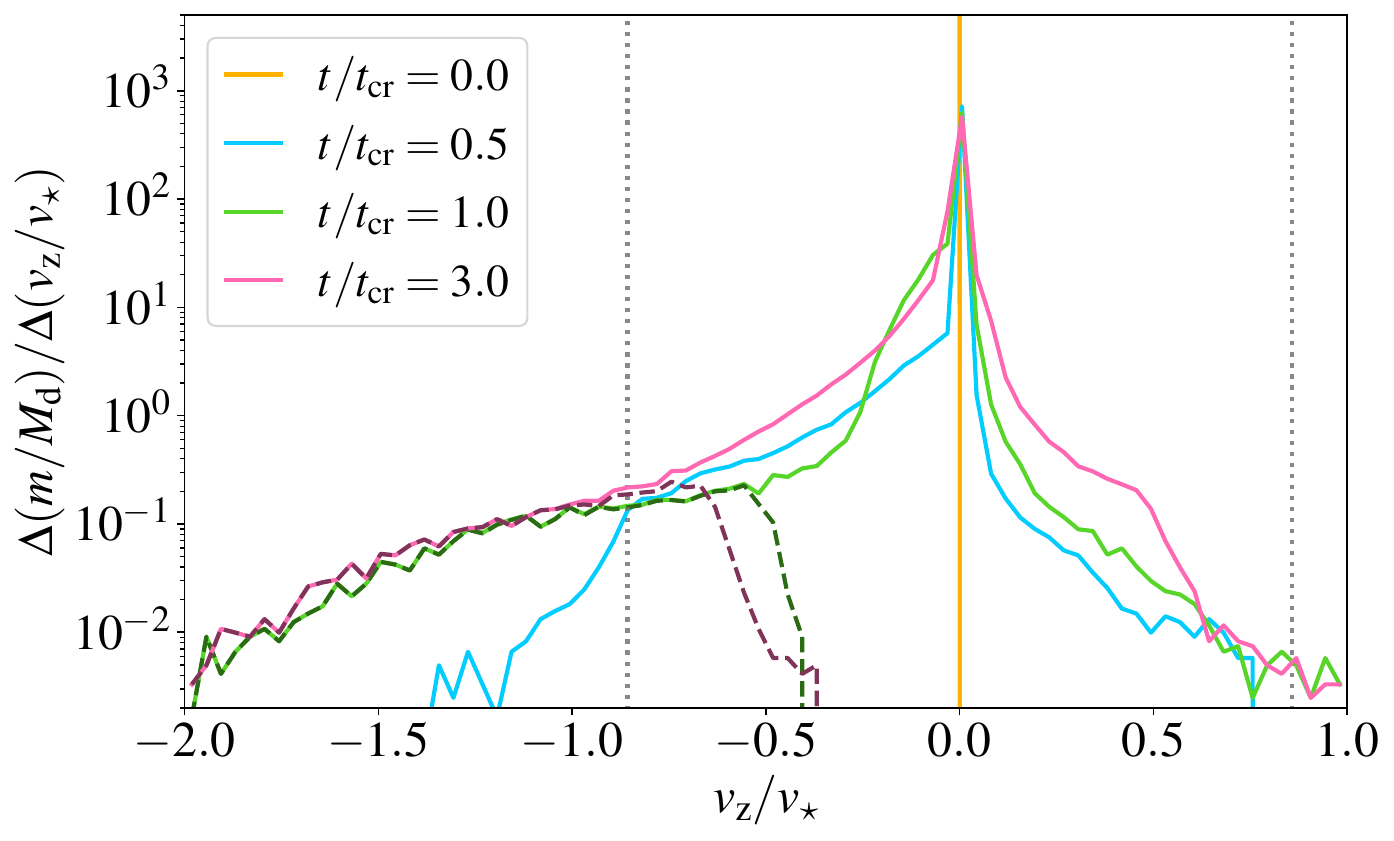}
\caption{Top: Momentum along the $z$-direction $P_\rm{z}^\rm{tot}$ of the gas that has been affected by the shock by the end of our simulation. Bottom: Mass-weighted distribution of $v_\rm{z}$ at different times for the shocked gas. The solid lines show distributions of all the shocked gas, while the dashed lines correspond to distributions of gas ejected during the shock breakout along the forward direction, which was selected as gas below the disc at $t/t_\rm{cr}=1$ (see dashed green rectangle in the right panel in Figure \ref{fig:qpe_sim_asymm} and the region denoted by dashed green line in Figure \ref{fig:vz_vs_r}). The vertical dotted grey lines indicate $\pm  v_\rm{z}/v_\star=\pm \frac{6}{7}$, which corresponds to the analytically estimated maximum post-shock velocity for a perpendicular collision. The vertical dash-dotted brown lines indicate the mean velocity $v_\rm{z}/v_\star=0.05$ at the disc edges at $t/t_\rm{cr}=3$.}
\label{fig:Pz}
\end{figure}

Figure \ref{fig:Pz} (top panel) shows the evolution of the vertical momentum $P_\rm{z}$ for gas that has been affected by the shock by the end of our simulation. We calculate it as the sum of momenta along $\boldsymbol{e}_\rm{z}$ of gas that is either outside the disc or with a specific internal energy $u>2u_\rm{d}$ in the last snapshot, where $u_\rm{d}$ is the value in the initial disc. While the star is inside the disc, it continuously injects momentum into the surrounding gas, and $P_\mathrm{z}$  decreases as gas is accelerated predominantly along $-\boldsymbol{e}_\mathrm{z}$. After the star exits the disc,  the momentum asymptotes to a constant value of $P_\mathrm{z}/(M_\rm{d} v_\star)\approx -0.45$, reflecting that the momentum of the shocked gas remains constant once the injection stops. The asymptotic value is lower than unity because gas colliding with the star at oblique angles (farther from the tip)  is elastically reflected along directions inclined from the vertical. Additionally, inelastic gas-gas interactions and shocks prevent a purely elastic redistribution of momentum.

The momentum redistribution is further examined in Figure \ref{fig:Pz} (bottom panel), showing the mass-weighted distributions of velocities $v_\mathrm{z}$ of shocked gas at different times, selected in the same way as in Figure \ref{fig:Pz} (top panel). At $t/t_\rm{cr}=0$ (orange line), all the gas is at rest with $v_\rm{z} = 0$  since the star has not yet started to cross the disc. When the star collides with the disc, it injects momentum into the surrounding gas, producing an asymmetric velocity distribution with a broader and more extended tail towards negative $v_\mathrm{z}$, since gas is accelerated predominantly in the direction of the stellar motion. At $t/t_\rm{cr}=0.5$ (cyan line), the star is still inside the disc, and the most negative gas velocity is given by the gas inside the shock cap with $v_\rm{z}/v_\star\approx -1$. Conversely, gas has already started to escape from the disc along the backward direction (positive velocities). The maximum velocity of this gas $\Delta v$ can be estimated using Rankine-Hugoniot conditions. Assuming perpendicular collisions and that all thermal energy $\Delta u$ is converted into kinetic energy, we obtain $\Delta v = \sqrt{2\Delta u}={2}{\left( \Gamma+1 \right)^{-1}} v_\star = \frac{6}{7}v_\star$ in agreement with the velocity spread in the simulation. 

At $t/t_\rm{cr}\approx 1.0$ (green lines), shortly after shock breakout occurs along the forward direction, a fraction of gas is ejected from the disc (dashed green line). This gas is also highlighted with a dashed green line in Figure \ref{fig:qpe_sim_asymm} (right panel) and in Figure \ref{fig:vz_vs_r}. This gas reaches velocities up to a maximum of approximately $v_\mathrm{z}/v_\star \approx -2$, consistent with the extreme theoretical limit for an idealized elastic collision.\footnote{This can be understood by considering an elastic collision between two bodies, where one body (the star) is rigid and significantly more massive. In the star's rest frame, gas colliding perpendicularly with the tip of the star approaches at $-v_\star$ and reflects with $+v_\star$, as dictated by conservation of momentum and energy. Transforming back to the disc frame gives a post-collision velocity of $2v_\star$, the maximum possible in the absence of pressure forces or dissipation.}. At later times (pink lines), the $v_\rm{z}$ distribution stabilizes, maintaining its maximum spread. The gas moving along the forward direction but not directly ejected during the breakout achieves a spread of $v_\rm{z}/v_\star\approx \Delta v/v_\star\sim 1$. In contrast, the backward-directed gas velocity distribution exhibits a steeper decline around $v_\rm{z}/v_\star \approx 0.5$. This occurs because the backward-ejected gas is typically farther from the symmetry axis before the collision than the forward-ejected gas (see Figure \ref{fig:vz_vs_r}), and thus, on average, experiences weaker energy injection. These velocity distributions reveal a clear asymmetry in both velocities and mass distribution, which is primarily driven by the directional nature of the initial momentum injection, resulting in a forward outflow with a higher momentum.

\begin{figure}
        \centering
    \includegraphics[width=0.99\linewidth]{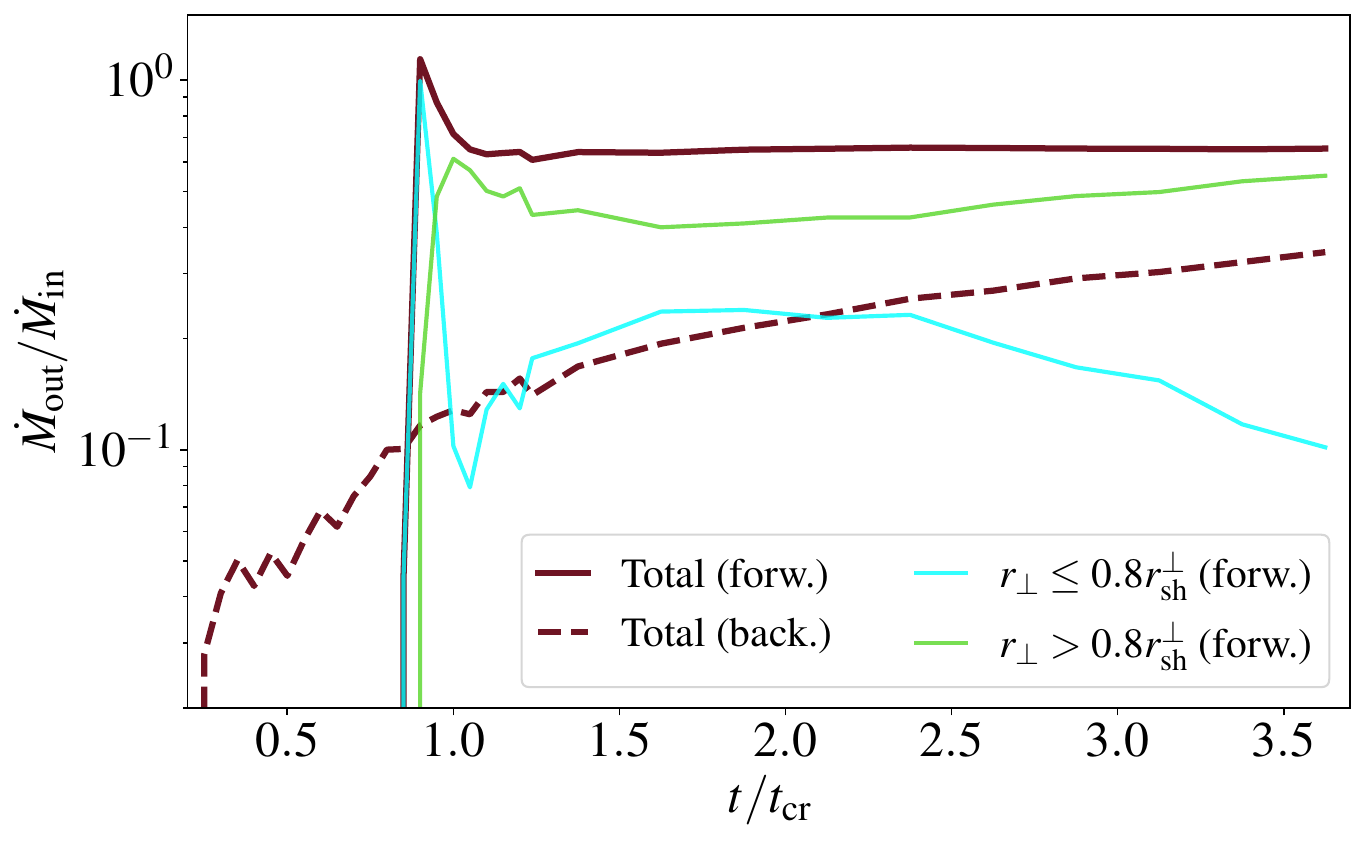}
    \caption{Mass outflow rate, $\dot{M}_\rm{out}$ , ejected along the forward (solid brown lines) and backward (dashed brown lines) direction. The cyan and green lines denote the gas, outflowing along the forward direction from inside and outside, respectively, of a cylindrical region with radius $r_\perp=0.8 r_\rm{sh}^\perp$. This region is illustrated in Figure \ref{fig:outflow_late} (see green region at the bottom disc surface), while the value of $r_\rm{sh}^\perp$ is denoted by an enlarged brown cross ($\boldsymbol{\times}$) symbol in Figure \ref{fig:r_of_t_shock}.}
        \label{fig:dotM_Mej}
\end{figure}

Figure \ref{fig:dotM_Mej} shows the mass outflow rate $\dot{M}_\rm{out}$ ejected along the forward (solid brown lines) and backward direction (dashed brown lines), normalized to the mass inflow rate into the shock $\dot{M}_\rm{in}= \pi R_\star^2 \rho v_\star = {1.7\times 10^{24}\, \mathrm{g/s}}$. The cyan and green lines represent two components of the forward outflow: cyan corresponds to gas ejected from within a cylinder of radius $0.8 r_\mathrm{sh}^\perp$, while green corresponds to gas ejected outside this region, where the lateral shock front distance {$r_\rm{sh}^\perp$ was calculated at $z/H=-1$ at different times in the same w}ay as in Figure \ref{fig:r_of_t_shock}. At early times, the gas contained in the shock cap is rapidly ejected during the forward-directed shock breakout, resulting in a sharp increase in $\dot{M}_\rm{out}$ (cyan lines). At later times ($t/t_\mathrm{cr} \gtrsim 2$), most of the forward outflow originates from the ring-like interface between the expanding shock and the disc edge (green line in Figure \ref{fig:dotM_Mej} and brown arrow lines in the left panel of Figure \ref{fig:outflow_late}), because the gas density in the cavity carved by the star decreases rapidly with time. For the backward outflow, $\dot{M}_\rm{out}$ is always dominated by the outflow near the interface between the shock front and the upper disc edge (see pink arrow lines in the left panel of Figure \ref{fig:outflow_late}).

The forward outflow exhibits higher $\dot{M}_\rm{out}$ than the backward one. At early times, this asymmetry is a consequence of the forward-directed shock breakout. At late times, the forward $\dot{M}_\mathrm{out}$ remains approximately constant, while the backward $\dot{M}_\mathrm{out}$ gradually increases and approaches the value of the forward $\dot{M}_\mathrm{out}$. The constant forward $\dot{M}_\mathrm{out}$ can be understood from the scaling of the mass outflow rate, $\dot{M}_\mathrm{out}\approx A\,\rho\,v_\mathrm{z}$, where $A$, $\rho$, and $v_\mathrm{z}$ are the effective outflow area, gas density, and vertical velocity of the gas crossing the ejection surface, respectively. At late times, most of the outflow is coming from a ring-like region of roughly fixed width, implying $A\propto r_\mathrm{sh}^\perp\propto t^{1/2}$ (see Figure~\ref{fig:r_of_t_shock}). From the jump conditions for a strong shock, $\rho$ remains an approximately constant ratio of the initial disc density. As expected from energy conservation,\footnote{Due to the laterally propagating shock, the specific thermal energy of the gas increases $ \propto v_{\mathrm{sh}}^{2}$. This energy then drives the expansion of the shocked gas and is converted into kinetic energy through radiation pressure. Since the post-shock flow is redirected primarily in the vertical direction, the resulting vertical velocity $v_z \propto v_{\mathrm{sh}}$.} we find that $v_\mathrm{z}\propto v_\mathrm{sh}\propto t^{-1/2}$, where $v_\mathrm{sh} = \mathrm{d} r_\mathrm{sh}^\perp / \mathrm{d}t$ is the shock front velocity. This leads to constant $\dot{M}_\mathrm{out}$ for the forward outflow. 

The situation is different for the backward direction, for which the outflow rate continuously rises. This is because, after passing through the shock, matter accumulates in the cavity before escaping in the outflow, with a time delay that is longer for gas deeper inside the disc. Over time, the outflow contains more of this accumulated matter, causing the backward $\dot{M}_\mathrm{out}$ to increase. To illustrate this behaviour, we show in Figure~\ref{fig:qpe_sim_asymm} the trajectory of two particles that escape at the same time, but with different time delays. The gas marked by the red '$\blacktriangle$' is shocked earlier and moves laterally within the disc before turning upwards, while the gas marked by the cyan '$\to$' is located closer to the disc edge and is shocked later but escapes more directly. Despite being shocked at different times, both particles emerge from the disc at $t/t_\mathrm{cr}\approx 1$. The overall effect on the backward outflow can be seen from Figure \ref{fig:vz_vs_r}, where orange and magenta contours show the gas that gets ejected within a fixed time interval at $t/t_\mathrm{cr}\approx 0.4$ and $t/t_\mathrm{cr}\approx 3.5$, respectively. As reflected by the size of the contours, more mass gets ejected over time, with the outflowing gas originating from deeper inside the disc and further away from the symmetry axis.

\subsection{Observable properties}\label{subsec:lcs}

\begin{figure}%[H] %  figure placement: here, top, bottom, or page
   \centering
        \includegraphics[width=0.49\textwidth]{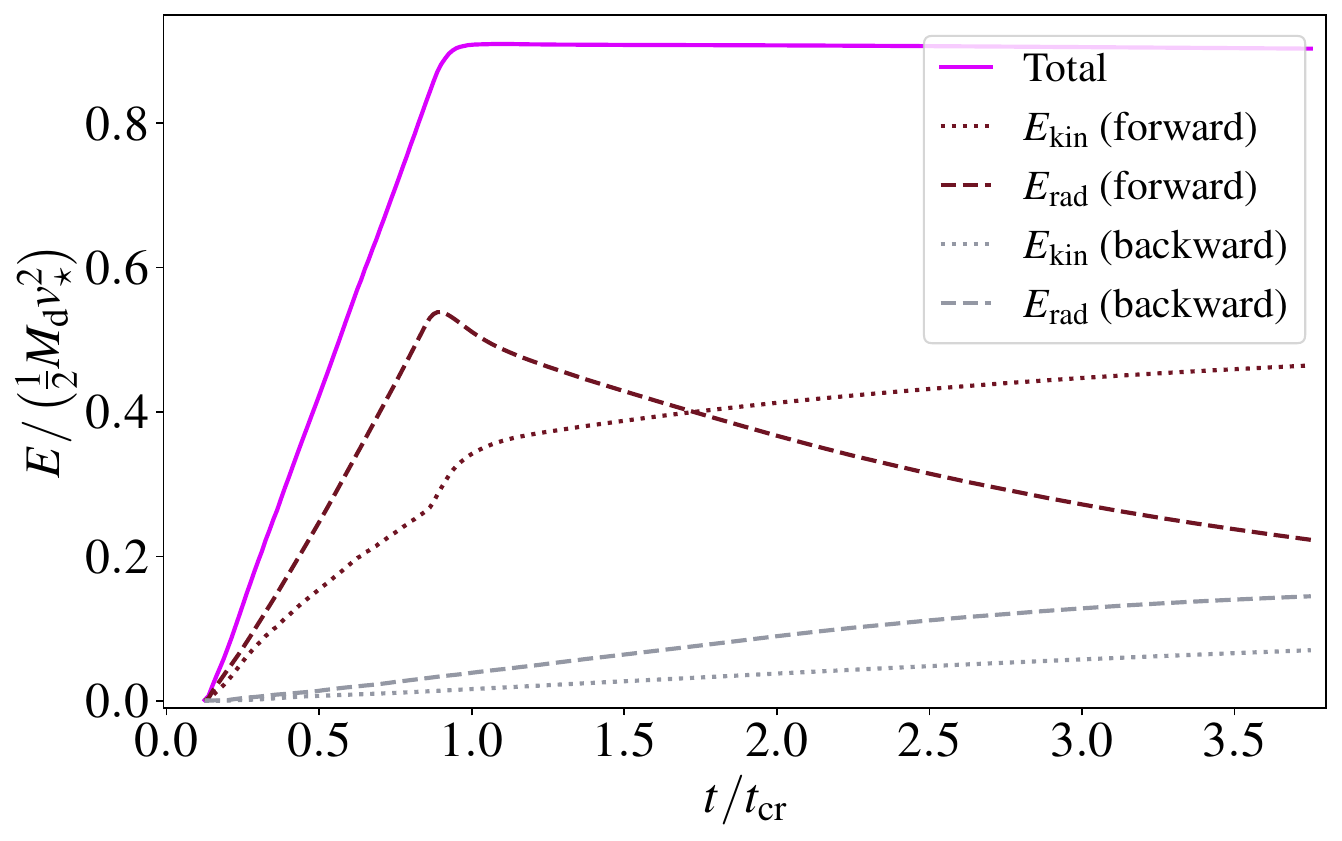}
   \caption{Total energy of all the gas $E_\rm{tot}$ (solid magenta line), $E_\rm{kin}$ of the forward- (dotted brown line) and backward-moving (dotted grey line) gas, $E_\rm{rad}$ inside the forward- (dashed brown line) and backward-moving (dashed grey line) gas.}
        \label{fig:Eplot}
\end{figure}

When a star collides with the disc, it injects energy which is first redistributed to thermal energy and then rapidly to kinetic energy of the gas $E_\mathrm{kin}$ and radiation energy $E_\mathrm{rad}$.  This process is shown in Figure \ref{fig:Eplot}, where we show different energy components for the gas moving forward (brown lines) and backward (grey lines). The values are normalized to the expected injected energy $\frac{1}{2}M_\rm{d} v_\star^2$.

While the star remains inside the disc, $E_\mathrm{rad}$ (dashed lines) steadily increases as new gas becomes shocked. Simultaneously, $E_\mathrm{kin}$ (dotted lines) grows, partially due to the conversion of $E_\mathrm{rad}$ to $E_\mathrm{kin}$ through radiation pressure in optically thick gas and partially due to direct mechanical pushing by the star. At $t/t_\rm{cr}\approx 0.9$, the star begins to exit the disc. As a result, new gas is no longer shocked, and $E_\mathrm{rad}$ of the forward-moving gas stops increasing. Simultaneously, $E_\mathrm{rad}$ in the forward outflow is rapidly converted into $E_\mathrm{kin}$ as the shocked gas expands into the surrounding vacuum, resulting in a prominent bump in $E_\mathrm{kin}$ at that time. For the backward-moving gas, $E_\mathrm{rad}$ continues to increase because a fraction of the downward-moving gas can reverse velocity (see trajectory of the brown '$\star$' particle in Figure \ref{fig:qpe_sim_asymm}). The total energy $E_\mathrm{tot}$ (solid magenta line) increases while the star is inside the disc and continues to inject energy. After the star exits the disc, $E_\mathrm{tot}/\left(\frac{1}{2}M_\rm{d}v_\star^2\right)$ asymptotes to $\approx 1$, since no additional energy is injected into the gas. We see that energy is not redistributed symmetrically, similar to momentum (see the bottom panel of Figure \ref{fig:Pz}). Specifically, the forward-moving gas carries more $E_\rm{rad}$ and $E_\rm{kin}$, mostly because the forward-moving gas contains more mass.\footnote{At $t/t_\rm{cr}=1$, both $E_\rm{rad}$ and $E_\rm{kin}$ of the forward-moving gas are approximately an order of magnitude larger than those of the backward-moving gas. This energy asymmetry is primarily due to the forward-moving component containing approximately ten times more mass, as inferred from the respective areas under the $v_\rm{z}<0$ and $v_\rm{z}>0$ distributions (pink lines) in Figure \ref{fig:Pz} (bottom panel). Although the forward-moving gas is, on average, faster and subjected to stronger shock heating (see Figure \ref{fig:vz_vs_r}), these effects are secondary.}

Radiation in the shocked outflows is transported by both diffusion and advection. To calculate the emerging luminosity $L$, we therefore include both contributions by integrating the sum of the advection flux $\boldsymbol{F}_\rm{adv}^\rm{ph}$ and the diffusion flux $\boldsymbol{F}_\rm{diff}^\rm{ph}$ across the photosphere.\footnote{To determine the photosphere, we considered a hemisphere of angular directions centred on the breakout point (the centre of the ejection region at the disc surface). Along each angular direction, we calculated the photospheric radius $R_{\mathrm{ph}}$ as the distance where the optical depth $\tau=\int_{R_{\mathrm{ph}}}^\infty \rho \kappa_{\mathrm{s}}\,\rm{d}r$ decreases to unity.} The diffusion flux $\boldsymbol{F}_\rm{diff}^\rm{ph}$ was calculated from Equation~\ref{eq:Fdiff}, while for the advective flux we followed \citet{Piro_2020ApJ...894....2P}, accounting for the motion of the photosphere, so that $\boldsymbol{F}_\rm{adv}^\rm{ph}=e_\rm{rad}(\boldsymbol{v}-\boldsymbol{v}_\rm{ph})$, with $\boldsymbol{v}$ and $\boldsymbol{v}_{\mathrm{ph}}$ the gas and photospheric velocities, respectively.\footnote{The photosphere velocity $\boldsymbol{v}_{\mathrm{ph}}=\Delta \boldsymbol{R}_{\mathrm{ph}}/\Delta t$ was determined from the change in the photosphere radius $\Delta \boldsymbol{R}_\rm{ph}$ between snapshots separated by a time $\Delta t$.} This approach ensured that only the net advected radiation energy was considered. Following \citet{Linial2023}, we estimate the peak luminosity as $ L_\rm{sym} = \frac{1}{2}\eta \dot{E}_\rm{in}= 2.1\times 10^{41}\, \mathrm{erg/s}$. It is obtained by assuming that the shock heating rate $\dot{E}_\rm{in}$ for the gas inside the column through which the star moves is reduced by a factor {$\eta=(V_\rm{f}/V_\rm{i})^{1/3}=0.07$} to account for adiabatic losses as this gas spherically expands from its initial volume $V_\rm{i}=(2\pi /7) R_\star^2 H$ to the final volume $V_\rm{f}=(4\pi/3) R_\rm{diff}^3$, where the radiation diffuses away. Here, $R_\rm{diff}=v_\star t_\rm{QPE}$ is the characteristic diffusion radius, and $t_\rm{QPE}=(\kappa_\rm{s} M_\rm{d})^{1/2}/(4\pi c v_\star)^{1/2}$ is the diffusion timescale. In the expression for $L_\rm{sym}$, the prefactor ${1}/{2}$  accounts for equal redistribution of energy into the two outflows.

\begin{figure}%[H] %  figure placement: here, top, bottom, or page
   \centering
    \includegraphics[width=0.49\textwidth]{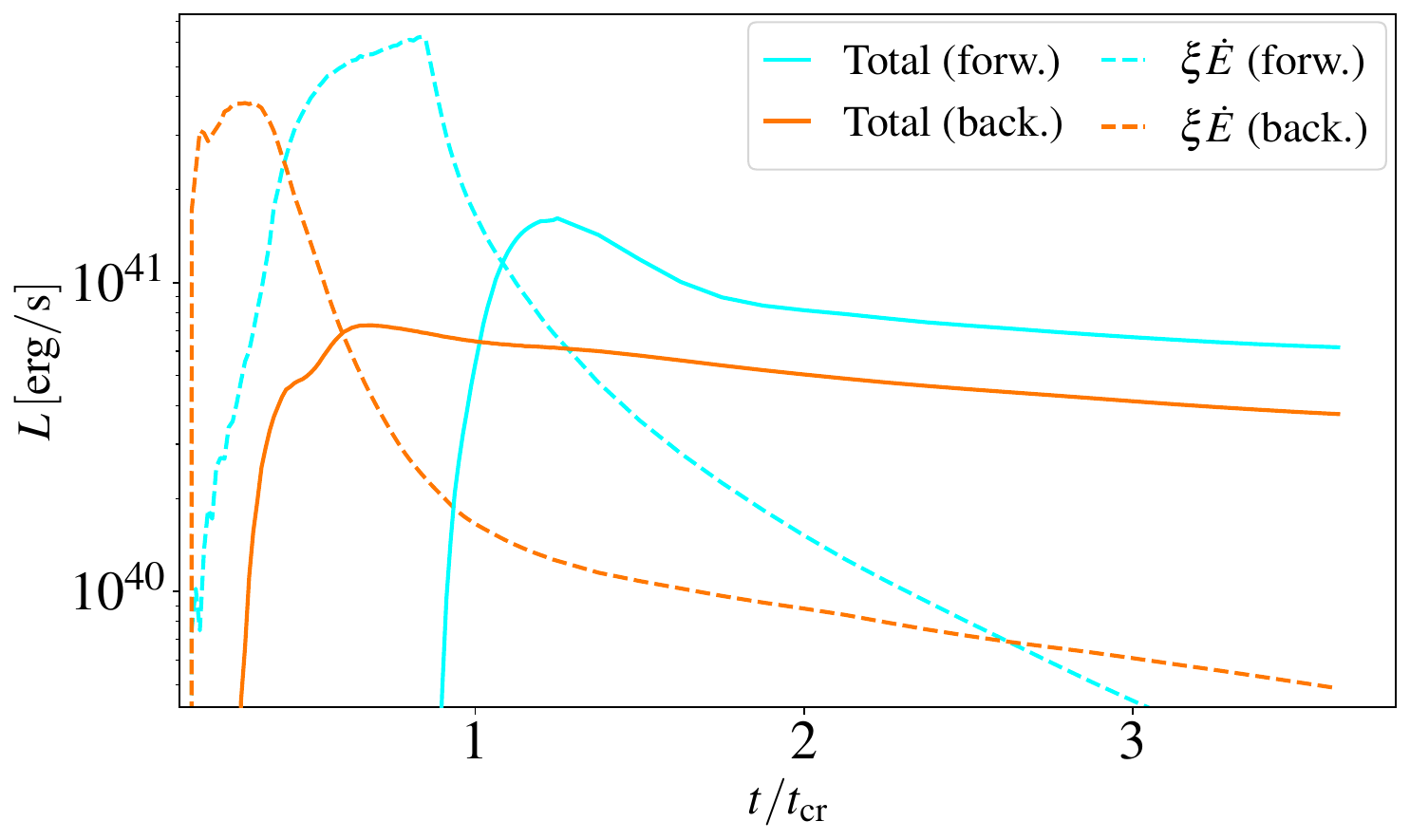}
    \includegraphics[width=0.49\textwidth]{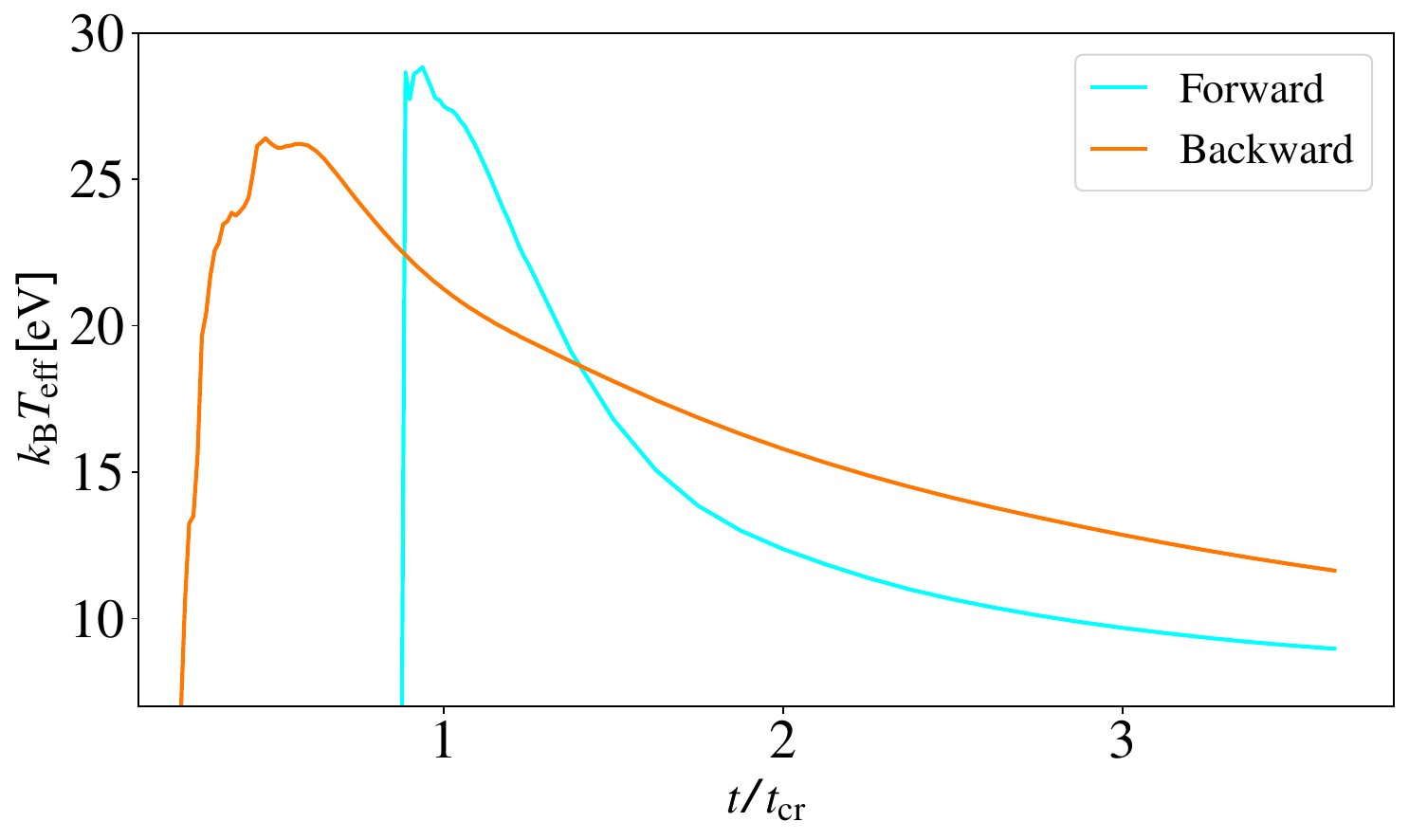}

   \caption{Top: Time evolution of $L$ of the forward (solid blue line) and backward (solid orange line) outflow. The dashed blue and orange lines denote $\dot{E}$ scaled by a factor $\xi=0.1$ for visual comparison, for the forward and backward outflows, respectively. To calculate $\dot{E}$ for each direction, we first identified particles with $v_\rm{z}<0$ (forward) and $v_\rm{z}>0$ (backward) in the final snapshot, and then tracked the $\dot{E}$ of these particles at all earlier times, summing them to obtain the total $\dot{E}$ of each outflow. {Bottom: Time evolution of $k_\rm{B}T_\rm{eff}$ of the forward (solid cyan line) and backward (dashed orange line) outflow.} }
        \label{fig:L_fiducial}
\end{figure}

Figure~\ref{fig:L_fiducial} (top panel) shows the time evolution of $L$ for the forward (solid cyan line) and backward (solid orange line) outflows. The light curves of both outflows show a rapid rise to $L \approx 10^{41}\,\mathrm{erg/s}\sim L_\mathrm{sym}$, followed by a more gradual decline. However, the forward outflow is about twice as luminous as the backward one. This asymmetry primarily results from the difference in $\dot{E}$ between the forward- and backward-moving gas, while the adiabatic losses are similar in both directions. We estimate the adiabatic losses from the ratio between the peak $L$ (solid lines in the top panel of Figure \ref{fig:L_fiducial}) and peak $\dot{E}$ (dashed lines in the top panel of Figure \ref{fig:L_fiducial}) for each outflow, obtaining a value of $\approx 0.02\sim \eta$ for both.\footnote{The limited role of adiabatic losses in producing the luminosity asymmetry can be inferred from the scaling of $\eta = (V_\mathrm{f}/V_\mathrm{i})^{1/3}$ with the ejecta mass $M_\mathrm{ej}$ and velocity $v_\mathrm{ej}$. The initial shocked volume scales as $V_\mathrm{i}\propto M_\mathrm{ej}$, because the gas is compressed to a post-shock density, which is independent of whether it is later ejected forwards or backwards. The final volume of the expanding ejecta scales as $V_\mathrm{f}\propto (v_\mathrm{ej} t_\mathrm{diff})^3$, where $t_\mathrm{diff}\propto M_\mathrm{ej}^{1/2} v_\mathrm{ej}^{-1/2}$ for a spherically expanding, optically thick outflow. Hence, $\eta \propto v_\mathrm{ej}^{1/2} M_\mathrm{ej}^{1/6}$. To produce a factor-of-two luminosity asymmetry, the forward outflow would therefore need to contain $\sim64$ times more mass or reach a velocity about four times higher than the backward outflow. Both values are larger than those obtained from our simulations, as inferred from the mass distributions of $v_\mathrm{z}$ (bottom panel of Figure~\ref{fig:Pz}).} Instead, the peak $\dot{E}$ of the forward-moving gas (dashed cyan lines) is about twice that of the backward one (dashed orange lines), consistent with the luminosity asymmetry. This difference arises because the backward-ejected gas is, on average, located farther from the symmetry axis (see Figure~\ref{fig:vz_vs_r}), which reduces the energy increase at the shock.

We quantify this effect analytically by comparing $\dot{E}$ for the two outflows in the star's rest frame. The shock heating rate can be written as $\dot{E} = \int_A \Delta u\,\rm{d}\dot{M}_\rm{in}$, where $\mathrm{d} \dot{M}_\mathrm{in}= \rho_\mathrm{d}  v_\rm{n}\,\mathrm{d}A$ is the mass inflow rate through a surface element $\mathrm{d}A = 2\pi R_\star^2 \sin\theta\,\mathrm{d}\theta$, $\theta$ is the angle measured from the symmetry axis, $\Delta u =0.37 v_\rm{n}^2$ is the thermal energy increase in a strong, radiation pressure-dominated shock,  and $v_\rm{n} = v_\star \cos\theta$ is the component of gas velocity normal to the stellar surface. For the forward-ejected gas, the integral is performed over $\theta \in [0, \pi/2]$. All of this gas is ejected during the forward shock breakout, since it does not have enough time to flow around the star (see the trajectory of the green '$\blacksquare$' particle and the gas inside the dashed green rectangle in the right panel of Figure~\ref{fig:qpe_sim_asymm}). For the backward-ejected gas, the situation is different, because gas located close to the tip of the star cannot be expelled backwards, as it requires more time to flow around the star (see the blue '$\blacklozenge$ trajectory in Figure~\ref{fig:qpe_sim_asymm}). As a result, only gas outside a conical region with $\theta>\theta_0$ contributes to the backward shock heating, where $\theta_0$ is the half-opening angle of the region separating forward- and backward-ejected material. We therefore evaluate the integral over $\theta\in[\theta_0,\pi/2]$, where $\theta_0 = 35^\circ$ is obtained from the simulation at $t/t_\mathrm{cr} = 0.3$, corresponding to the time of the backward $\dot{E}$ peak (dashed orange line in the top panel of Figure \ref{fig:L_fiducial}). {By evaluating the integral for $\dot{E}$ over these surfaces, we obtain a forward-to-backward ratio of $\dot{E}$ equal to $\cos^{-4}\theta_0 \approx 2.2$, consistent with the asymmetry in the peak $\dot{E}$ obtained from the simulations.}

The forward light curve (solid cyan line in the top panel of Figure \ref{fig:L_fiducial}) features a break to a slower decline at $t/t_\rm{cr}\approx1.8$. We interpret this break as a shift in the dominant source of escaping radiation: the peak is associated with radiation emerging from gas ejected in the forward shock breakout, while the later part of the light curve is powered by radiation produced by shocked gas escaping the disc after spending a longer time in the cavity. We verified this by selecting the gas elements that contribute to the luminosity at different times and tracing their trajectories backwards.\footnote{To identify the gas contributing to the luminosity at each time, we selected particles located near the trapping surface, since outside this surface the photons decouple from the gas by diffusing away, and the luminosity is roughly constant. The trapping surface was determined as the distances along each angular direction in the hemisphere centred on the breakout point, where the optical depth reaches $\tau = c/v$,  where $v$ is the local gas velocity.} For $t/t_\mathrm{cr}\lesssim 1.8$, this material can be traced back to the forward breakout ejecta, whereas gas reaching the trapping surface at later times originates from the cavity.

{In Figure~\ref{fig:L_fiducial} (bottom panel) we show the evolution of $k_\mathrm{B}T_\mathrm{eff}$ for the forward (solid cyan line) and backward (dashed orange line) outflows, where $T_\mathrm{eff}$ is the effective temperature and  $k_\mathrm{B}$ is the Boltzmann constant. We compute $T_\mathrm{eff}$ from the Stefan--Boltzmann law
$T_\rm{eff}^4 = L/(S_\rm{ph}\sigma)$, where $\sigma$ is the Stefan--Boltzmann constant. We obtained $L$ and $S_\mathrm{ph}$ from simulations as the bolometric luminosity (solid cyan and solid orange lines in the top panel of Figure~\ref{fig:L_fiducial} for the forward and backward outflows, respectively) and the photospheric surface area, respectively, of the corresponding outflow. We find that the forward outflow reaches a peak temperature $k_\mathrm{B}T_\mathrm{eff}\approx 29$~eV, which is slightly higher than for the backward outflow. Additionally, the forward outflow cools more rapidly due to its faster expansion. For both outflows, $T_\mathrm{eff}$ starts to decrease before $L$ reaches its peak value, because $S_\mathrm{ph}$ increases rapidly during the initial expansion phase immediately after breakout. The peak temperatures are also within a factor of two of the analytic estimate of the blackbody temperature
$k_\mathrm{B}T_\mathrm{BB} \approx k_\mathrm{B}(c u_\gamma/4\sigma)^{1/4}\approx 16\,\mathrm{eV}$, where $u_\gamma = L_\mathrm{sym}\tau/(4\pi R_\mathrm{diff}^2 c)$ is the radiation energy density and $\tau = c/v_\mathrm{ej}$ is the optical depth at $R_\mathrm{diff}$ (e.g. \citealt{Linial2023}).}

\section{Discussion} \label{sec:discussion}

\subsection{Effect of key system parameters and comparison with other works}\label{subsec:effect_params}

The dynamics of star-disc collisions and the properties of the resulting outflows depend on key system parameters such as intercepted disc mass $M_\rm{d}$, disc vertical extent $H$, stellar radius $R_\star$, stellar velocity $v_\star$, stellar orbit inclination w.r.t. the midplane of the disc $i$, vertical disc density profile (i.e. not uniform, but Gaussian), and temperature of the unperturbed disc. Although our current study focuses on a specific (fiducial) set of parameters, we briefly discuss here how variations in these parameters could qualitatively affect the characteristics of the outflows and their observable signatures compared to the fiducial case. We defer a detailed quantitative parameter exploration to future work.

i) Increasing $M_\rm{d}$ would increase the injected momentum and energy. However, the velocity of the shocked gas would remain largely unchanged, as it is set by the elastic collision limit or by post-shock acceleration due to radiation pressure, where the available specific energy budget is determined by the characteristic increase in the specific internal energy of the gas due to the collision---both of which are independent of $M_\rm{d}$ (see Section \ref{subsec:outflow} for more details). As a result, the additional injected momentum and energy would be redistributed into proportionally more massive outflows moving at similar velocities. The hydrodynamic evolution, including the outflow geometry and shock structure, is therefore expected to be similar to the fiducial case. However, the increased outflow mass would lead to higher optical depths and likely different light curves. In particular, the escape of radiation would likely be delayed, potentially broadening and delaying the peaks in luminosity.

ii) Increasing $v_\star$ would increase injected momentum and energy,  while keeping the shocked mass unchanged.  Due to this, the outflows would expand more rapidly. However, both the forward and backward outflow would expand faster by a similar factor, so we expect the gas dynamics to remain comparable to the fiducial case when evaluated at the same $t/t_\rm{cr}$. Nonetheless, we expect that increasing $v_\star$ would affect the light curves due to the larger shock heating rate. In particular, this would likely produce brighter and narrower luminosity peaks.

iii) For the same values of $R_\star/H$ (while keeping $M_\rm{d}$ fixed), the gas dynamics of star-disc collisions is expected to remain largely unchanged due to approximate scale invariance, as suggested by \citet{Yao_2025ApJ...978...91Y}. However, deviations may arise for different $R_\star/H$ because the mass expelled during the forward shock breakout scales as $M_\rm{cap}\propto M_\rm{d} \left(R_\star/H\right)$ (see Equation \ref{eq:Mcap}), while the masses of outflows scale approximately as $\propto M_\rm{d}$. For constant $M_\rm{d}$ , this implies that increasing $R_\star/H$ likely leads to larger $M_\rm{cap}$, and therefore to more pronounced asymmetry in the masses and luminosities of the forward and backward outflows.

iv) Decreasing the inclination angle $i$ would break the symmetry of the interaction, leading to a less axisymmetric outflow geometry. Additionally, for lower $i$, the star's path length through the disc would be longer, causing more gas to be shocked. This would lead to larger injected momentum and energy, resulting in more massive and energetic outflows, expanding with similar velocities as in the fiducial case. This is similar to the effect of decreasing $R_\star/H$, while keeping $R_\star$ and $M_\rm{d}$ constant. {We therefore expect the forward and backward outflows to become more similar in luminosity, while the light curves may become broader than in the fiducial case due to the increased outflow mass and energy.} Additionally, higher inclination increases the duration over which disc layers at different radial distances from the SMBH are shocked, which may increase the effect of radial shear between adjacent flow layers. We discuss possible consequences of this in Section \ref{sec:limit}.

v)  For a disc with a Gaussian vertical density profile, but the same surface density, the total injected momentum and energy, as well as the velocity of the shocked gas, remain approximately the same as in the fiducial case. However, the light curves will likely be different. In the case of a Gaussian disc, the shock heating rate is higher in the denser layers near the disc midplane, while the outer, less dense layers receive weaker heating. This produces a stronger gradient in radiation energy density and hence a larger diffusion flux (see Equation \ref{eq:Fdiff}) at the photosphere. As a result, both outflows are expected to reach higher peak luminosities than in the fiducial case.

vi) The characteristic midplane temperature of an accretion disc involved in star-disc collisions is expected to be $k_\rm{B}T\sim 40\,$eV (e.g. \citealt{Linial2023, Yao_2025ApJ...978...91Y, vurm_2025ApJ...983...40V}), corresponding to a specific thermal energy $u_\rm{real}/v_\star^2\sim 10^{-2} $ for a radiation-pressure dominated gas. This is significantly higher than the value $u_\rm{d}/v_\star^2\sim 10^{-5}$ assigned to our unperturbed disc (see Section \ref{subsec:star_disc}). However, this choice is unlikely to affect the early-time shock dynamics, as the characteristic post-shock thermal energy increase is $\Delta u/v_\star^2\approx 0.37$, which greatly exceeds both $u_\rm{d}$ and plausible $u_\rm{real}$ values. At later times, however, a higher $u_\rm{real}$ could cause the shock to stall sooner, as the shock velocity drops below the local sound speed earlier. This would modify the late-time dynamics and may shorten the observable flare duration.

{We note that these expectations are broadly consistent with the results of \citet{Huang_2025ApJ...993..186H} and \citet{vurm_2025ApJ...983...40V}, particularly with respect to parameters i), ii), and iii). Recent hydrodynamical studies by \citet{Liu_2026arXiv260300226L} and \citet{Huang_2026arXiv260400953H} likewise support the importance of collision geometry and the resulting asymmetry between the outflow, especially in relation to parameter iv).} However, a direct comparison or extrapolation is not straightforward, as these works differ in several key aspects. \citet{Huang_2025ApJ...993..186H} performed 2D radiation-hydrodynamic simulations using a fiducial setup that includes a Gaussian disc. They also use a multi-group (frequency-dependent) radiation treatment, whereas we adopt a grey (frequency-integrated) flux-limited diffusion approximation. On the other hand, \citet{vurm_2025ApJ...983...40V} performed 1D Monte Carlo radiation transport simulations focusing on scenarios in which the gas does not flow around the star, which differs from our setup.
{  \citet{Liu_2026arXiv260300226L} performed global 3D hydrodynamical simulations including the gravity of the SMBH and disc rotation, and found that star-disc collisions generate strongly asymmetric forward and backward ejecta. \citet{Huang_2026arXiv260400953H} performed 2D and 3D hydrodynamical simulations in a local frame and showed that oblique collisions due to disc rotation can significantly enhance the backward ejecta and reduce the luminosity asymmetry, although the forward component is still brighter.}

\subsection{Comparison to observations}\label{sec_disc:observations}

Observed QPE flares typically reach peak X-ray luminosities of approximately $10^{41}-10^{43}\,\rm{erg\,s^{-1}}$ and have durations on the order of hours (e.g. \citealt{Miniutti2019, Arcodia2021, Arcodia2022}) or even days \citep{Hernandez_2025NatAs...9..895H}. Some sources, including GSN 069 \citep{Miniutti2023} and eRO-QPE2 \citep{Arcodia_2024A&A...690A..80A}, exhibit an alternating 'strong-weak' luminosity pattern, where the peak luminosity of consecutive flares differs by between a few percent and a factor of two. For our simulation, the peak bolometric luminosity of the outflows is $\sim 10^{41}\,\rm{erg\,s^{-1}}$ (see Figure \ref{fig:L_fiducial}), placing it on the lower end of the observed range. Furthermore, the peak $L$ of the forward outflow is approximately two times higher than the peak $L$ of the backward outflow, while at later times $L$ of the forward outflow is $\approx 50\%$ higher than $L$ of the backward outflow. While we do not capture the entire duration of the flare, we can extrapolate the light curve to later times. This gives timescales of approximately an hour when $L$ decreases by an order of magnitude from the peak. These results suggest that the star-disc collision scenario can reproduce not only the observed peak luminosities and timescales of QPEs, but also the alternating 'strong-weak' flare pattern seen in specific sources. In addition, if an accretion disc is viewed close to edge-on, radiation from both the forward and backward outflows could be detected during a single collision. The resulting light curve could show the brighter outflow as the main peak, while the less luminous outflow may appear as a secondary peak or plateau. This effect could account for the more complex flare morphologies observed in some QPEs, such as the double-peaked or plateau-like structures observed in eRO-QPE1 \citep{Arcodia2022}. {However, we note that the star-disc collision model, in the regime where the star remains largely unperturbed by the interaction, may not explain all QPE sources. In particular, such collisions likely do not generate enough energy to reproduce the radiated energies of some long-recurrence systems, such as AT2019qiz \citep{Nicholl_2024Natur.634..804N,Linial_2025ApJ...991..147L,Mummery_2025arXiv250421456M}.}

Light curves and the asymmetry can be further influenced by the system parameters (see Section \ref{subsec:effect_params}). {Additionally, the luminosity asymmetry between the forward and backward outflows may be influenced by the fact that QPEs are typically detected in a narrow soft X-ray band. To estimate this effect, we used the effective temperature as a qualitative proxy for spectral hardness (see the bottom panel of Figure~12). The resulting temperature differences imply that even if the bolometric luminosities were comparable, their soft X-ray luminosities could differ, potentially enhancing or reducing the luminosity asymmetry. However, we note that the peak effective temperatures $\sim 30$~eV, obtained from our simulation, are significantly lower than the blackbody temperatures inferred from observed QPE flares of $\sim 100$--$200$~eV (e.g. \citealt{Kara_2025ARA&A..63..379K}).} This discrepancy reflects our use of the LTE assumption, which may break down in QPE-like conditions that are photon-starved (see, e.g. \citealt{Linial2023, vurm_2025ApJ...983...40V, Huang_2025ApJ...993..186H}). We discuss these limitations further in Section \ref{sec:limit}. A realistic prediction of the SED and its observational asymmetry will require non-LTE modelling, which is beyond the scope of this study.

\subsection{Limitations and future improvements}\label{sec:limit}

In our simulation, the star was modelled as a rigid sphere interacting via elastic collisions with SPH particles representing the disc gas. This idealized setup allowed us to isolate and study the fundamental processes governing shock formation and asymmetric outflow formation, without introducing additional complexities associated with stellar structure. {In particular, we do not resolve the star's internal radial density profile, which determines how deeply the ram pressure penetrates into the stellar envelope, how much energy is absorbed by the star, and how much mass is stripped during the collision. In reality, the stellar atmosphere is compressible and dynamically responds to the ram pressure exerted during the collision.} As a result, the interactions are no longer perfectly elastic, since a fraction of the injected energy is absorbed by the stellar envelope, leading to puffing, deformation, and possibly mass stripping, as shown by \citet{Yao_2025ApJ...978...91Y}. This energy loss would reduce the amount of energy available to power the outflows, likely leading to slower expansion and lower luminosities. We aim to explore these effects in future work.

{A further limitation is that we neglect the momentum exchange between the star and the disc. While this exchange is negligible for the single collision studied here, over many repeated passages the cumulative momentum exchange with the disc, together with mass loss through ablation (see paragraph above), is expected to modify the stellar orbit, changing the disc-crossing location and flare recurrence properties. Repeated collisions would likely drive the orbit towards a nearly coplanar configuration (e.g. \citealt{Subr_1999A&A...352..452S}), at which point the repeated-collision flare phase would end.}

{From a numerical point of view, the SPH nature of the \textsc{Phantom} code could introduce additional limitations, since thin shocks, contact discontinuities, and small-scale wake instabilities are generally less sharply resolved than in high-order grid-based codes.  \citet{Huang_2026arXiv260400953H} showed that accurate shock capturing in star--disc collisions mainly requires resolving the thin bow shock and its stand-off distance. In our case, the stand-off distance is $\sim 0.1R_\star$, consistent with the expected high-Mach rigid-body regime and supporting the conclusion that the bow shock is resolved. \citet{Prust_2024MNRAS.527.2869P} showed that for wake instabilities to develop, a sufficiently extended low-Mach region behind the obstacle --- i.e. a separation bubble --- must exist and persist long enough for perturbations to grow rather than be advected away. In the weakly gravitating regime, where the Bondi radius of the colliding object satisfies $R_{\rm B}\ll R_\star$, the trailing flow is characterized by a small separation bubble with almost no wake instabilities, whereas in more strongly gravitating cases the separation bubble is larger and the wake turbulence is stronger. In our setup, $R_{\rm B}\sim 10^{-4}R_\star$, placing the interaction in the weakly gravitating regime, so we expect the effect of wake turbulence to be negligible.}

The radiation transfer method implemented in \textsc{Phantom} uses a flux-limited diffusion approximation, which assumes LTE and isotropic radiation fields. However, the LTE assumption may break down in star-disc collisions due to inefficient photon production and rapid expansion of the shocked gas. In such cases, the thermalization timescale can exceed the dynamical timescale, leading to photon starvation—a regime in which the photon number density is too low to maintain thermal equilibrium (see e.g. \citealt{vurm_2025ApJ...983...40V}; \citealt{Linial2023}). {Photon starvation is expected to be more easily achieved in the forward outflow during shock breakout, where the dynamical timescale is shorter than in the backward outflow, making thermalization less efficient \citep[e.g.][]{vurm_2025ApJ...983...40V,Huang_2025ApJ...993..186H}.} This may further enhance the observed luminosity asymmetry between forward and backward outflows, especially at early times. As the system evolves and the expansion slows, the severity of photon starvation is expected to diminish, reducing its influence on the luminosity asymmetry. The emerging spectrum can also be affected by Compton up-scattering of photons by free electrons. \citet{Huang_2025ApJ...993..186H} and \citet{vurm_2025ApJ...983...40V} find that Comptonization leads to a harder spectrum than predicted under LTE assumptions, bringing the modelled spectral energy distributions into closer agreement with the observed soft X-ray SEDs of QPEs.

\section{Conclusions} \label{sec:conclusion}

Collisions between a star and an accretion disc are one of the most promising models to explain QPEs. In this study, we performed the first 3D radiation-hydrodynamics simulation of a star-disc collision, {focusing on the regime where} the star remains unperturbed by the collision and the stellar crossing time through the disc is sufficiently long for shocked gas to flow around the star. We focused on the dynamics of the collision, the properties of the resulting outflows, and their observable radiation signatures. We also investigated the physical origin of the asymmetry between the forward and backward outflows and assessed whether this asymmetry can reproduce the 'strong-weak' flare pattern observed in several QPEs. Our main conclusions are as follows.

(i) A star crossing an accretion disc at supersonic speed drives a quasi-paraboloidal bow shock and injects momentum preferentially along its direction of motion. This intrinsically anisotropic injection, together with lateral shock expansion and the formation of a secondary shock in the stellar wake, leads to an asymmetric redistribution of energy and momentum. As a result, two outflows emerge on opposite sides of the disc with different gas trajectories and geometries, momenta and energies, and luminosities.

(ii) The forward outflow consists of fast ejecta produced during the quasi-spherical shock breakout of gas directly in front of the star when it exits the disc, and of gas shocked and pushed aside by the star that either flows around it and is later redirected outwards by the secondary shock, or expands laterally before escaping from the forward disc edge. In contrast, the backward outflow does not undergo a quasi-spherical breakout. As a result, the forward outflow is more extended and closer to spherical.

(iii) The forward-ejected gas carries more momentum and energy, and it experiences a higher shock heating rate. This asymmetry arises because the star injects momentum predominantly along its direction of motion. Moreover, the gas expelled forwards typically originates closer to the tip of the star, where the shocks are strongest, compared to the backward-ejected gas.

(iv) The emerging bolometric luminosity of the forward outflow is about twice that of the backward outflow, with a more pronounced early peak produced by the shock breakout on the forward side. This luminosity asymmetry mainly reflects the stronger shock heating of forward-ejected gas as described in iii).

(v) The luminosity asymmetry between the forward and backward outflows is consistent with the 'strong-weak' pattern observed in several QPE sources. This suggests that asymmetric injection and redistribution of energy and momentum during star-disc collisions can naturally account for the observed luminosity asymmetry.

This study provides a physically motivated framework for exploring star-disc collisions as a potential origin of QPEs. The asymmetry in outflow properties and luminosity arises naturally from the collision dynamics, offering a possible explanation for the alternating 'strong-weak' flare patterns seen in several QPE sources. Future work will extend the parameter space, incorporate more realistic stellar structures, and compare simulated light curves to observations. These efforts will help test the viability of the star-disc collision scenario and could offer a new avenue to constrain the properties of SMBHs, the formation and structure of accretion flows, and the stellar dynamics in galactic centres.

\section*{Data availability}

The scripts used to construct a local section of an accretion disc and to simulate star-disc collisions will be made publicly available as a part of \textsc{Phantom} available at \href{https://github.com/danieljprice/phantom}{https://github.com/danieljprice/phantom}. The software used to analyse simulation snapshots is available at \href{https://github.com/tajjankovic/Radiation-hydrodynamics-of-star-disc-collisions/}{https://github.com/tajjankovic/Radiation-hydrodynamics-of-star-disc-collisions/}. \footnote{Movies made from the simulations are available online at \href{https://www.youtube.com/playlist?list=PLH8qhWjKWQ92nPx_tPaPYnobRCPUjlfdF}{https://www.youtube.com/playlist?list=PLH8qhWjKWQ92nPx\_tPaPYnobRCPUjlfdF}.}
\begin{acknowledgements}
Researcher T. J. conducts his research under the Marie Skłodowska-Curie Actions – COFUND project, which is co-funded by the European Union (Physics for Future – Grant Agreement No. 101081515). T. J. and A.J. acknowledge the financial support from the Slovenian Research Agency (research core funding P1-0031, infrastructure program I0-0033, and project grants Nos. J1-8136, J1-2460, N1-0344). T. J. acknowledges the use of HPC cluster Phoebe of the Central European Institute of Cosmology (CEICO) at the Institute of Physics of the Czech Academy of Sciences where the computations were performed. Funded by the European Union (ERC, Unleash-TDEs, project number 101163093). Views and opinions expressed are however those of the author(s) only and do not necessarily reflect those of the European Union or the European Research Council. Neither the European Union nor the granting authority can be held responsible for them. This work was co-funded by the European Union and supported by the Czech Ministry of Education, Youth and Sports (MEYS) (Project No. CZ.02.01.01/00/22\_008/0004632 -- FORTE). We thank X. Huang for the useful discussions.

The following software was used in this work: 
     \textit{Matplotlib} \citep{Hunter:2007},
       \textit{NumPy} \citep{harris2020array}, 
    \textit{SciPy} \citep{2020SciPy-NMeth}.
    
\end{acknowledgements}

\bibliographystyle{aa}
\bibliography{bibliography}

\begin{appendix}

\section{Accumulation of radiation}\label{app:rad_accum}

A potential concern in our simulations is the absence of a surrounding low-density medium that would allow radiation to escape freely once it reaches the outer boundary of the outflows. This could lead to artificial radiation accumulation in the outer, optically thin regions and potentially propagate this effect inwards to the optically thick regime, affecting the gas dynamics and introducing deviations from physically realistic behaviour. In this section, we derive the expected scalings for the radiation energy density \( e_\mathrm{rad} \)  across different radiative regimes and compare them to simulation results in order to assess whether significant radiation buildup occurs in our simulation.

We consider three different radiative transport regimes, defined by the radial coordinate $r$ measured outwards from the breakout point (the centre of the corresponding ejection region at the surface of the disc): the optically thick regime ($r \leq R_\mathrm{tr}$), the diffusion regime ($R_\mathrm{tr} \leq r \leq R_\mathrm{ph}$), and the free-streaming regime ($r \geq R_\mathrm{ph}$), where $R_\mathrm{tr}$ is the trapping radius and $R_\mathrm{ph}$ is the photospheric radius.

In the optically thick regime, radiation is tightly coupled to the gas, and the flow is dominated by radiation pressure. Since radiation pressure scales as $P_{\mathrm{rad}} \propto \rho^{4/3}$, and $P_{\mathrm{rad}} \propto e_{\mathrm{rad}}$, the expected scaling is
\begin{equation}\label{eq:thick}
e_{\mathrm{rad}} \propto \rho^{4/3}.
\end{equation}
In the diffusion regime, radiative luminosity is approximately conserved with $L \approx  4\pi r^2 F \approx \text{const.}$, where $F \approx e_{\mathrm{rad}} c / \tau$ is radiative flux and  $\tau \approx \rho \kappa_\rm{s} r$ is optical depth. The scaling is then
\begin{equation}\label{eq:diff}
e_{\mathrm{rad}} \propto \frac{\rho}{r}.
\end{equation}
In the free-streaming regime, radiation escapes freely and $F \approx e_{\mathrm{rad}} c$. Assuming constant $L$, the scaling is
\begin{equation}\label{eq:free}
e_{\mathrm{rad}} \propto r^{-2}.
\end{equation}

\begin{figure}
        \centering
\includegraphics[width=\linewidth]{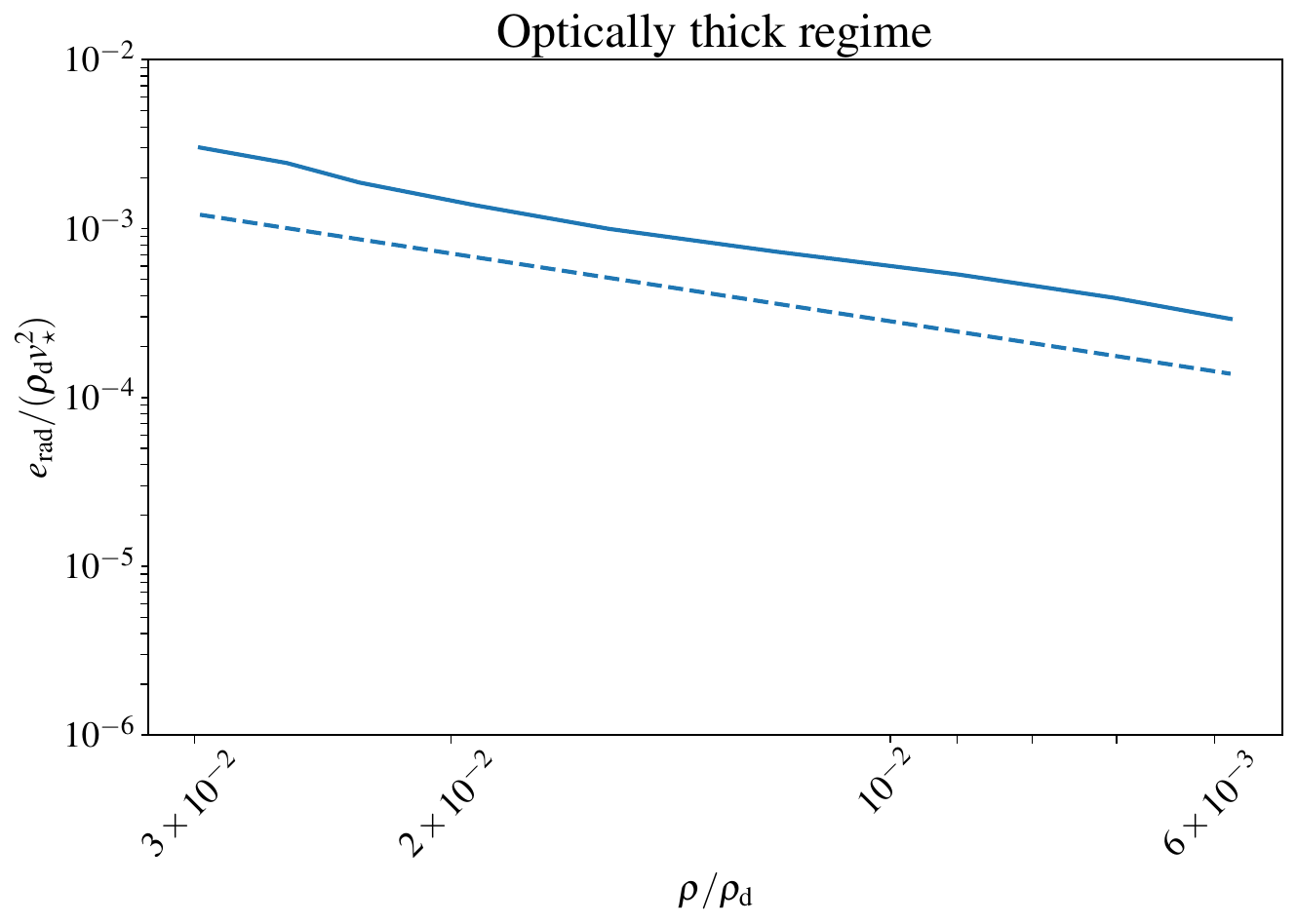}
\par\medskip
  \includegraphics[width=\linewidth]{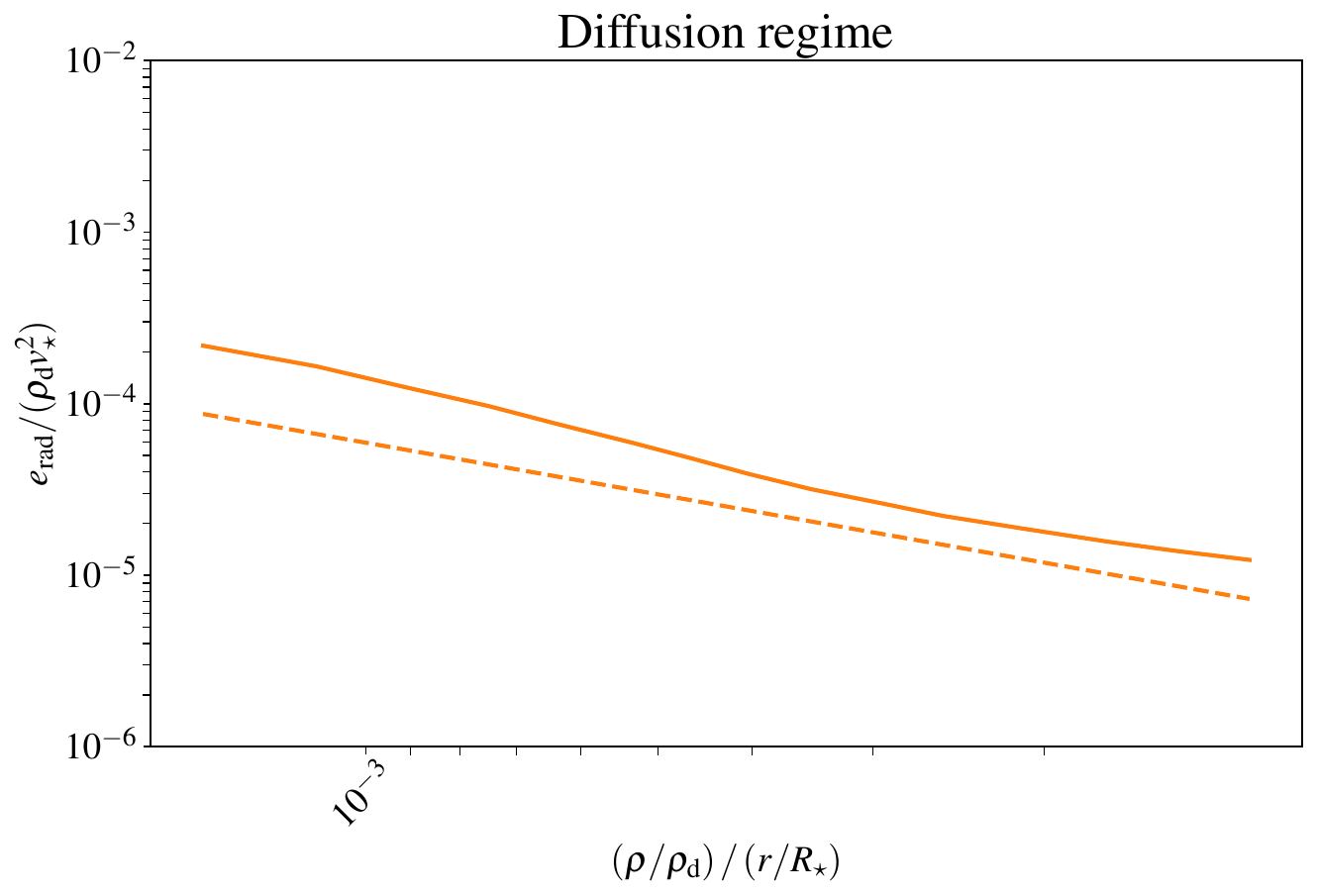}
 \par\medskip
  \includegraphics[width=\linewidth]{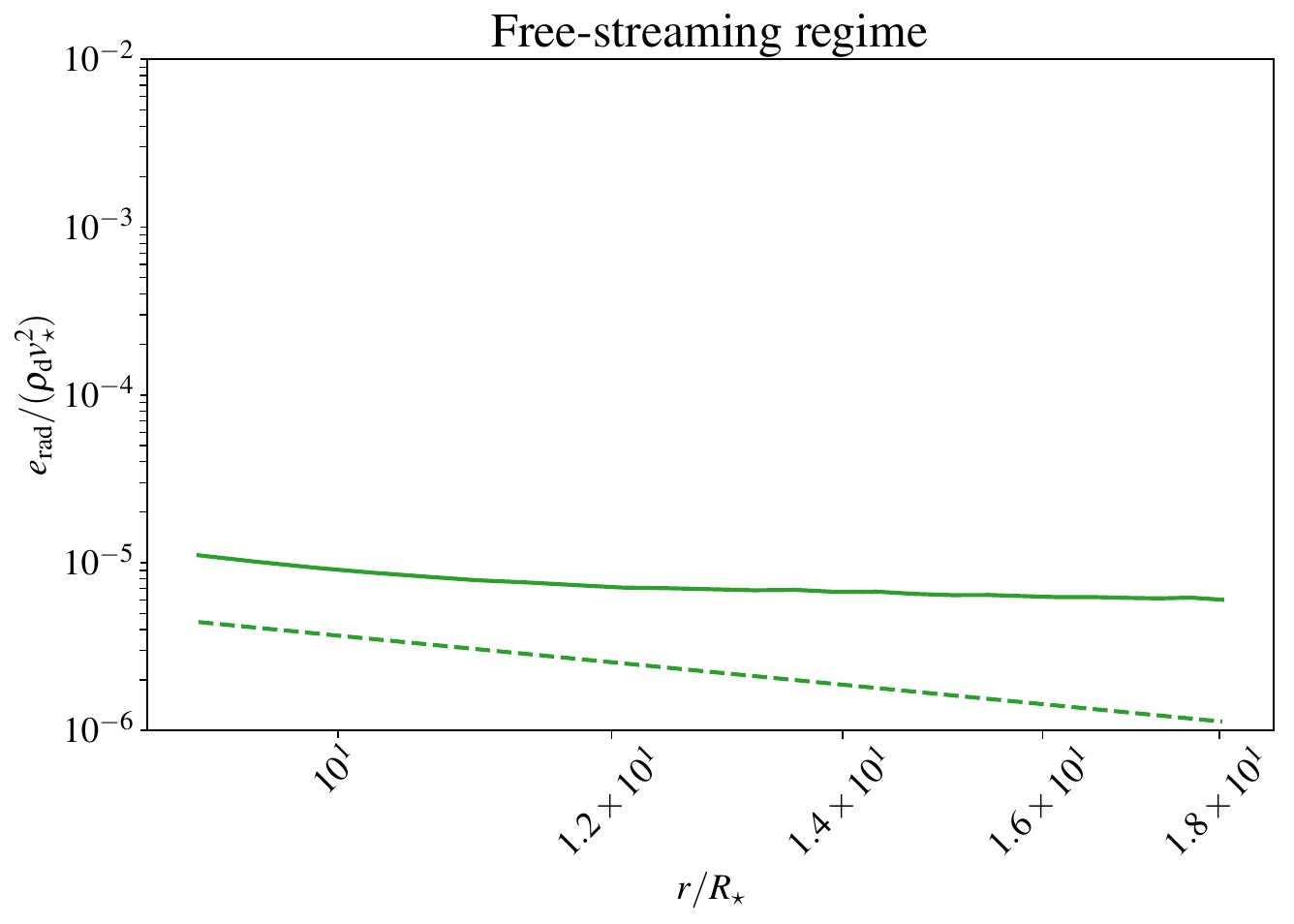}
        \caption{Radial profiles of radiation energy density $e_{\mathrm{rad}}$ for the forward outflow at $t/t_\rm{cr}=2.5$ in spherical shells. Simulation results (solid lines) are compared to analytic predictions (dashed lines) in the optically thick (top), diffusion (middle), and free-streaming regimes (bottom).}
        \label{fig:erad_qpe}
 \end{figure}

Figure \ref{fig:erad_qpe} compares radial profiles of $e_\rm{rad}$ obtained from simulations at $t/t_\rm{cr}=2.5$ (solid lines) to analytic scalings (dashed lines)  derived in Equations \ref{eq:thick}, \ref{eq:diff}, and \ref{eq:free}. The simulation results agree well with the predicted behaviour across the optically thick and diffusion regimes. At large radii, in the free-streaming regime, modest deviations appear due to numerical radiation accumulation near the outer simulation boundary. These results suggest that the radiative transfer implementation in \textsc{Phantom} captures the expected physical scalings across the relevant transport regimes. The accumulation effects are confined to the outermost regions and do not significantly influence the interior gas dynamics.

\newpage

\section{Quasi-steady structure of the shock cap}\label{app:Mcap}

We consider cases in which the vertical thickness of the disc is several times larger than $R_\star$. In this regime, the mass inflow rate into the bow shock $\dot{M}_\mathrm{in}$ is balanced by the outflow rate of gas that escapes around the sides of the star $\dot{M}_\mathrm{out}$. This balance establishes a quasi-steady state in which both the radial thickness of the shock front $\Delta R$, and the total mass enclosed within the shock cap $M_\rm{cap}$ remain approximately constant. In the following, we provide analytic estimates for $\Delta R$ and $M_\rm{cap}$, with the derivations carried out in the rest frame of the star.

The mass inflow rate into the shock cap can be estimated as
\begin{equation}
\dot{M}_\mathrm{in} \approx \pi R_\star^2 \rho_\rm{d} v_\star.
\end{equation}
The mass outflow rate is approximately given by
\begin{equation}
\dot{M}_\mathrm{out} \approx 2\pi R_\star \Delta R \rho_\mathrm{sh} v_\mathrm{out},
\end{equation}
where $\rho_\mathrm{sh}=\rho_\rm{d}({\Gamma + 1})({\Gamma - 1})^{-1}$ is the density of the shocked gas and $v_\mathrm{out}$ is the velocity at which shocked gas flows along the sides of the star. To estimate $v_\rm{out}$, we apply Bernoulli's equation along a streamline extending from the stagnation point at the shock tip (where the velocity is approximately zero) to the escape point at the side of the star. Assuming negligible gas pressure at the escape point, Bernoulli's equation gives $P_\mathrm{in}{\Gamma}({\Gamma - 1})^{-1}  \approx 2^{-1}\rho_\mathrm{sh} v_\mathrm{out}^2$, where the stagnation pressure is given by post-shock condition $P_\mathrm{in} = {2 \Gamma}({\Gamma +1})^{-1} \rho_\rm{d} v_\star^2$. This yields $v_\rm{out}\approx 2\Gamma (\Gamma+1)^{-1}v_\star$.

Assuming a quasi-steady state, the inflow and outflow rates are approximately equal,
\begin{equation}\label{eq:dotM_equil}
\dot{M}_\mathrm{in} \approx \dot{M}_\mathrm{out}, 
\end{equation}
which gives
\begin{equation}
\Delta R \approx \frac{R_\star}{2} \frac{\rho_\rm{d}}{\rho_\mathrm{sh}} \frac{v_\star}{v_\mathrm{out}} \approx\frac{\Gamma-1}{4\Gamma}R_\star\approx {0.1}\, \frac{R_\star}{R_\odot}
\end{equation}
and 
\begin{flalign}
\nonumber M_\mathrm{cap} &\approx 2\pi R_\star^2 \Delta R \rho_\mathrm{sh} \approx \frac{\pi}{2}\frac{\Gamma+1}{\Gamma}R_\star^3 \rho_\rm{d}  && 
\label{eq:Mcap}\\
    &\approx 1.7\times 10^{-8} M_\odot \left(\frac{R_\star}{R_\odot}\right)^3 \left(\frac{\rho_\rm{d}}{3.7\times 10^{-8}\,\mathrm{g\,cm^{-3}}}\right)&&\\
    &\approx 0.1 M_\rm{d} \left(\frac{R_\star}{R_\odot}\right) \left(\frac{H}{3R_\odot}\right)^{-1},&&
\end{flalign}
where we assumed $\Gamma=4/3$ and $M_\rm{d}=2\pi R_\star^2H \rho_\rm{d}$. We compared this analytic estimate of $M_\rm{cap}$ to the mass of the forward outflow in the simulation at $t/t_\rm{cr}=1$, when half of the star has exited the disc (see panels of Figures \ref{fig:qpe_sim_density} and \ref{fig:qpe_sim_asymm} and early peak in $\dot{M}_\rm{out}$ in Figure \ref{fig:dotM_Mej}). The analytic result matches the simulation value to within $10\%$.

\end{appendix}

\end{document}